\documentclass[12pt,a4paper,twoside]{report}
\usepackage[applemac]{inputenc}
\usepackage{amssymb}
\usepackage{amsmath}
\usepackage{amsopn}
\newcommand{\LL}{\mathbb{L}}
\newcommand{\CC}{\mathbb{C}}

\newcommand{\KK}{\mathbb{K}}

\newcommand{\NN}{\mathbb{N}}

\newcommand{\RR}{\mathbb{R}}
\newcommand{\ZZ}{\mathbb{Z}}

\newcommand{\frA}{\mathfrak{A}}
\newcommand{\fram}{\frA(M)}

\newcommand{\frb}{\mathfrak{B}}

\newcommand{\frp}{\mathfrak{p}}

\newcommand{\kbm}{\mathcal{B}(M)}

\newcommand{\ktm}{\mathcal{T}(M)}
\newcommand{\ktrm}{\mathcal{T}_{r}(M)}

\newcommand{\kD}{\mathcal{D}}

\newcommand{\kf}{\mathcal{F}}

\newcommand{\kh}{\mathcal{H}}
\newcommand{\kI}{\mathcal{I}}
\newcommand{\kj}{\mathcal{J}}

\newcommand{\kL}{\mathcal{L}}
\newcommand{\mm}{\mathcal{M}}

\newcommand{\kP}{\mathcal{P}}
\newcommand{\kQ}{\mathcal{Q}}
\newcommand{\kR}{\mathcal{R}}
\newcommand{\kS}{\mathcal{S}}

\newcommand{\kT}{\mathcal{T}}

\newcommand{\ga}{\alpha}

\newcommand{\gd}{\delta}
\newcommand{\eps}{\varepsilon}

\newcommand{\gG}{\Gamma}

\newcommand{\gk}{\kappa}
\newcommand{\gl}{\lambda}

\newcommand{\gO}{\Omega}
\newcommand{\gf}{\varphi}

\newcommand{\gs}{\sigma}

\newcommand{\tm}{\subseteq}

\newcommand{\∞}{\infty}

\newtheorem{definition}{Definition}[chapter]
\newtheorem{proposition}{Proposition}[chapter]
\newtheorem{theorem}{Theorem}[chapter]
\newtheorem{lemma}{Lemma}[chapter]
\newtheorem{corollary}{Corollary}[chapter]
\newtheorem{remark}{Remark}[chapter]
\newtheorem{example}{Example}[chapter]

\newtheorem{discussion}{Discussion}[chapter]

\newcommand{\por}{\kP_{0}(\kR)}

\newcommand{\pr}{\kP(\kR)}

\newcommand{\qr}{\mathcal{Q}(\mathcal{R})}
\newcommand{\qm}{\kQ(\kT(M))}
\newcommand{\qptm}{\kQ^{pt}(\kT(M))}
\newcommand{\qrm}{\kQ(\ktrm)}
\newcommand{\qxm}{\kQ_{x}(\ktm)}
\newcommand{\qxrm}{\kQ_{x}(\ktrm)}

\newcommand{\qmm}{\mathcal{Q}(\mm)}
\newcommand{\qA}{\mathcal{Q}(\frA)}

\newcommand{\ql}{\mathcal{Q}(\LL)}

\newcommand{\lh}{\mathcal{L}(\mathcal{H})}

\newcommand{\all}{\forall}
\newcommand{\ex}{\exists}
\newcommand{\rr}{\kR}

\newcommand{\hr}{\kR_{sa}}
\newcommand{\el}{E_{\gl}}

\newcommand{\emm}{E_{\mu}}

\newcommand{\egf}{E^{\gf}}
\newcommand{\egfl}{E^{\gf}_{\gl}}

\newcommand{\eal}{E^{A}_{\gl}}

\newcommand{\ea}{E^{A}}

\newcommand{\ef}{E^{f}}
\newcommand{\efl}{E^{f}_{\gl}}

\newcommand{\sfl}{\gs_{f}(\gl)}

\newcommand{\we}{\wedge}
\newcommand{\We}{\bigwedge}
\newcommand{\Ve}{\bigvee}

\newcommand{\tto}{\mapsto}
\newcommand{\lra}{\Longrightarrow}

\newcommand{\smm}{\setminus}

\newcommand{\dr}{\kD(\rr)}

\newcommand{\pp}{\perp}
\newcommand{\irr}{\in \RR}

\newcommand{\mir}{\mu \in \RR}
\newcommand{\lir}{\gl \in \RR}

\newcommand{\nin}{n \in \NN}

\newcommand{\kik}{k \in \KK}

\newcommand{\urb}{\overset{-1}}
\newcommand{\ffi}{\kf(\kI)} 
\newcommand{\qAm}{\kQ(\frA(M))}

\begin{document}

\title{\Huge{Observables \\ III : Classical Observables}}

\author{Hans F.\ de Groote\footnote{degroote@math.uni-frankfurt.de;
FB Mathematik, J.W.Goethe-Universität Frankfurt a.\ M.}}
\date{5.1.2006}
\titlepage
\maketitle
\begin{abstract}
    In the second part of our work on observables we have shown that
    quantum observables in the sense of von Neumann, i.e.
    \emph{bounded} selfadjoint operators in some von Neumann subalgebra
    $R$ of $L(H)$, can be represented as bounded continuous functions 
    on the Stone spectrum $Q(R)$ of $R$. Moreover, we have shown that 
    this representation is linear if and only if $R$ is abelian, and
    that in this case it coincides with the Gelfand transformation of 
    $R$. In this part we discuss classical observables, i.e. measurable
    and continuous functions, under the same point of view. We obtain 
    results that are quite similar to the quantum case, thus showing
    up the common structural features of quantum and classical observables.
\end{abstract}

\begin{center}
    {\large Für Karin}
\end{center}

\tableofcontents

\chapter{Introduction and Overview}
\label{in}
\pagestyle{myheadings}
\markboth{Introduction}{and Overview}

\begin{quote}
    \emph{Man soll \"{o}fters dasjenige untersuchen, was von den
    Menschen meist vergessen wird, wo sie nicht hinsehen und was so
    sehr als bekannt angenommen wird, dass es keiner Untersuchung mehr
    wert geachtet wird.}\\
    (Georg Christoph Lichtenberg)
\end{quote}
~\\
~\\
A classical observable is a function on phase space which is,
depending on the context, measurable, continuous or smooth. We will
consider here measurable and continuous observables in the spirit of
quantum observables. A quantum observable is a selfadjoint operator
$A$ (densely defined) on some Hilbert space $\kh$. The physical
meaning of $A$, however, becomes manifest in the spectral family $\ea
= (\el)_{\lir}$ of $A$. The spectral theorem makes precise how the
operator can be recovered from its spectral family. Moreover, we have 
seen that each bounded observable $A$, contained in a von Neumann
subalgebra $\rr$ of $\lh$, induces a bounded continuous function
$f_{A} : \qr \to \RR$ on the Stone spectrum $\qr$ of the projection
lattice $\pr$ of $\rr$. The mapping $f^{\rr}_{\ast} : \hr \to
C_{b}(\qr, \RR), \quad A \tto f_{A}$ is a generalization of the Gelfand
transformation: it is linear if and only if $\rr$ is abelian, and in this case
$f^{\rr}_{\ast}$ is (essentially) the Gelfand transformation for $\rr$. 
We recall from part II (\cite{deg4}) that $f_{A}$ is defined via the spectral
family $\ea$ of $A$:
\[
    \all \ \frb \in \qr : \ f_{A}(\frb) := \inf \{ \lir \mid \eal \in 
    \frb \}. 
\]
Since this construction works for spectral families in an arbitrary
(countably) complete lattice, it can be applied also to measurable and
to continuous functions - provided that these classes of functions can
be characterized by spectral families in suitable lattices\footnote{It
is precisely this point why we do not include smooth functions in our 
discussion: we simply do not know good conditions on the spectral families
of continuous functions on a smooth manifold that characterize smooth 
functions. We think that this is an interesting and promising open
problem.}. \\
~\\
In section \ref{coa} we show that there is a natural bijection between
the set $\fram(M, \RR)$ of $\fram$-measurable functions
$\gf : M \to \RR$, where $\fram$ is a sub-$\gs$-algebra of the $\gs$-algebra
$pot(M)$ of all subsets of the set $M$, and the set $\kS(\fram)$ of all
spectral families in $\fram$. The bijection is given by
\[
    \begin{array}{cccc}
        \gf^{\fram} : & \kS(\fram) & \to & \fram(M, \RR)  \\
         & E & \tto & \gf_{E},
    \end{array}
\]    
where $\gf_{E} : M \to \RR$ is defined as
\[
    \all \ x \in M : \ \gf_{E}(x) := \inf \{ \lir \mid x \in \el \}.
\]
If $\frA$ is an abstract $\gs$-algebra, it is therefore natural to
regard a spectral family $E : \RR \to \frA$ as a generalized
$\frA$-measurable function. \\
~\\
We mentioned already in part II (\cite{deg4}) that the notion of observable
function can be generalized to spectral families in countably complete
orthomodular lattices. Let $\LL$ be a countably complete orthomodular 
lattice and let $E$ be a bounded spectral family in $\LL$. The function
$f_{E} : \ql \to \RR$, defined by
\[
    \all \ \frb \in \ql : \ f_{E}(\frb) := \inf \{ \lir \mid \el \in
    \frb \},
\]
is called the (observable) function associated to $E$. We prove that
the mapping $f_{\ast} : E \tto f_{E}$ is injective and that $f_{E}$ is
a bounded continuous function on $\ql$. \\
~\\
Let $\frA$ be an abstract $\gs$-algebra. According to the theorem of Loomis
and Sikorski (\cite{Sik}), $\frA$ can be represented as a quotient $\fram / \kI$
of a sub-$\gs$-algebra $\fram \tm pot(M)$ modulo a suitable $\gs$-ideal
$\kI$ in $\fram$. It was shown in \cite{deg3} that the Stone spectrum 
$\qA$ of $\frA$ is homeomorphic to the Gelfand spectrum of the abelian
$C^{\ast}$-algebra $\kf_{\fram}(M, \CC) / \kf(\kI)$, where $\kf_{\fram}(M,
\CC)$ denotes the $C^{\ast}$-algebra of all \emph{bounded}
$\fram$-measurable functions $M \to \CC$ and $\kf(\kI) \tm \kf_{\fram}(M,
\CC)$ the closed selfadjoint ideal of all $\gf \in \kf_{\fram}(M,
\CC)$ that vanish outside some $A \in \kI$. Let $\gf \in
\kf_{\fram}(M, \RR)$, $\egf$ the spectral family of $\gf$ and let
$f_{\gf} := f_{\egf}$. Then a minor modification of the proof of
theorem 2.9 in \cite{deg4} shows that, up to the homeomorphism $\qAm
\simeq \gO(\kf_{\fram}(M, \CC))$, the mapping $\kf_{\fram}(M, \RR) \to
C(\qAm, \RR), \ \ \gf \tto f_{\egf},$ is the restriction of the Gelfand
transformation to $\kf_{\fram}(M, \RR)$. Let $\kf_{\RR}(\kI) := \kf_{\fram}(M, \RR)
\cap \ffi$. Using results from \cite{deg3}, we show that the mapping
$\gf \tto f_{\egf}$ induces a mapping
\[
    \begin{array}{cccc}
      \gG : & \kf_{\fram}(M, \RR) / \kf_{\RR}(\kI) & \to & C(\qA, \RR)  \\
       & [\gf] & \tto & f_{[\egf]},
    \end{array}
\]
where $[\gf] := \gf + \kf_{\RR}(\kI)$ and $[\egf]$ denotes the
equivalence class of $\egf$ modulo $\kI$, and that $\gG$ is the
restriction of the Gelfand transformation $\gG : \kf_{\fram}(M, \CC) / \kf(\kI)
\to C(\qA)$ to the selfadjoint part $\kf_{\fram}(M, \RR) / \kf_{\RR}(\kI)$ 
of $\kf_{\fram}(M, \CC) / \kf(\kI)$. As a corollary, we obtain that
each bounded spectral family $E$ in $\frA$ is the quotient $[\egf]$ of some
bounded spectral family $\egf$ in $\fram$ modulo $\kI$. Therefore, a
bounded generalized measurable function can be regarded as an equivalence
class of an ordinary bounded measurable function $M \to \RR$.\\
~\\
In section \ref{cob} we discuss continuous functions on a Hausdorff
space $M$ under the same point of view as for measurable functions. We show
that the natural assignment of a spectral family $\ef$ in $\ktm$ to a continuous
function $f : M \to \RR$ is given by
\[
     \efl := int \urb{f}(]-\∞, \gl]).
\]
The continuity of $f$ implies that the spectral family $\ef$ has the
property
\[
  (\ast) \qquad   \all \ \lir \ \all \ \mu > \gl : \ \overline{\efl} \tm \ef_{\mu},
\]    
hence $\ef$ is a spectral family in the complete Boolean lattice
$\ktrm$ of all regular open subsets of $M$. \\
Conversely, if a spectral family $E$ in $\ktm$ is given, 
\[
    f_{E}(x) := \inf \{ \lir \mid x \in \el \}
\]
is a real number only if $x$ is an element of
\[
      \kD(E) := \{x \in M \mid \exists \ \gl \in \RR : \ x \notin \el) \}.
\] 
In general, $\kD(E)$ is different from $M$. This is due to the
definition of the meet of an infinite family in $\ktm$. Of course,
$\kD(E) = M$ if $E$ is bounded from below. In any case, $\kD(E)$ is a 
dense subset of $M$. Thus $E$ induces a function $f_{E} : \kD(E) \to
\RR$. The function $f_{E}$ is continuous if and only if the spectral
family $E$ is \emph{strongly regular}, i.e. satisfies the condition
$(\ast)$ above. The interplay between spectral families and continuous
functions is completely described in theorem \ref{co33}.\\
~\\
Now let $\gf : M \to \RR$ be a \emph{bounded} continuous function with
spectral family $\egf$. As for measurable functions, we define a
continuous function $f_{\egf} : \qm \to \RR$ by
\[
    f_{\egf}(\frb) := \inf \{ \lir \mid \egfl \in \frb \}.
\]
In this way we obtain a mapping 
\[
    \begin{array}{cccc}
        f_{\ast} : & C_{b}(M, \RR) & \to & C(\qm, \RR)  \\
         & \gf & \tto & f_{\egf}.
    \end{array}
\]
In general, $f_{\ast}$ is not surjective. To determine the range of
$f_{\ast}$, we need the following notion. If $x \in M$, a quasipoint
$\frb \in \qm$ is called a \emph{quasipoint over $x$} if $x \in
\overline{U}$ for all $U \in \frb$. Since $M$ is a Hausdorff space, a 
quasipoint $\frb \in \qm$ is a quasipoint over at most one $x \in M$. 
We denote by $\qxm$ the set of all quasipoints over $x$. Let
\[
    \qptm := \bigcup_{x \in M}\qxm. 
\]   
Then $\qptm$ is dense in $\qm$, and we obtain a surjective mapping
$pt : \qptm \to M$, defined by $\all \ x \in M : \ pt(\qxm) := \{x\}$.
We show that $pt$ is continuous and identifying. \\
Now we can describe the range of $f_{\ast}$ by the following property:
$f \in C(\qm, \RR)$ is in the range of $f_{\ast}$ if and only if the
restriction of $f$ to $\qptm$ (which determines $f$ uniquely) factors 
over $pt$. This is equivalent to the property that $f$ is constant on 
each fibre of $pt$.\\ It is easy to generalize these results to complex 
spectral families and complex valued functions and we discuss bounded 
complex spectral families and their associated functions in a more
general context.\\
Let $C^{pt}(\qm)$ be the set of all $f \in C(\qm)$ that are constant
on each fibre of $pt$. It is obvious that $C^{pt}(\qm)$ is a
$C^{\ast}$-subalgebra of $C(\qm)$. We prove that $C_{b}(M)$ is
$\ast$-isomorphic to $C^{pt}(\qm)$ and that a canonical isomorphism is
given by 
\[
    \begin{array}{cccc}
	f_{\ast} : & C_{b}(M) & \to & C^{pt}(\qm)  \\
	 & \gf & \tto & f_{E^{\gf}},
    \end{array}
\]
where $E^{\gf}$ is the complex spectral family corresponding to
$\gf$ (theorem \ref{co43}).\\
~\\
Moreover, we show that the uniform approximation of measurable and
continuous functions by step functions can be seen as spectral
representations.\\
~\\
In the last section, we discuss the common features of quantum and
classical observables which are now apparent from our theory.
    

\chapter{Classical Observables}
\label{chap: CO}

In the previous part (\cite{deg4}) we have seen that selfadjoint elements 
$A$ of a von Neumann algebra $\rr$ correspond to certain continuous
real valued functions $f_{A}$ on the Stone spectrum (\cite{deg3})
$\qr$ of $\pr$.\\
~\\  
In the present part we will show that continuous real valued 
functions on a topological space $M$ can be described by a 
spectral families with values in the complete lattice $\kT(M)$ of 
open subsets of $M$. These spectral families $E : \RR \to \kT(M)$
can be characterized abstractly by a certain property of 
the mapping $E$. Thus also a classical observable has a ``quantum
mechanical'' description. In this part of our work, we will exhibit
further common features of quantum and classical observables. Similar
results hold for functions on a set $M$ that are measurable with respect
to a $\gs$-algebra of subsets of $M$. We start with the case of measurable
functions because it is technically simpler.\\
~\\  
We remind the reader of the definition of a spectral family in a general
complete lattice $\LL$:

\begin{definition}\label{co1}
  Let $\LL$ be a complete lattice. A spectral family in $\LL$ is a 
  mapping $E : \RR \to \LL$ with the following properties:
  \begin{enumerate}
       \item  [(1)] $\el \leq \emm$ for $\gl \leq \mu$,
       
      \item  [(2)] $\el = \bigwedge_{\mu > \gl}\emm$ for all $\gl 
		    \in \RR$, and
  
      \item  [(3)] $\bigwedge_{\gl \in \RR}\el = 0, \ 
		    \bigvee_{\gl \in \RR}\el = 1$.
  \end{enumerate}
  Note that this definition also applies to $\aleph_{0}$-complete
  lattices (usually called $\gs$-complete lattices), for the countable
  set of rationals is dense in $\RR$.
\end{definition}

\section{Measurable Functions}
  \label{coa}
\pagestyle{myheadings}
\markboth{Classical Observables}{Measurable Functions}  

Let $\fram$ is a sub-$\gs$-algebra of the power set $pot(M)$ of 
some non-empty set $M$ (that means that $\bigwedge_{n}U_{n} = 
\bigcap_{n}U_{n}, \ \Ve_{n}U_{n} = \bigcup_{n}U_{n}$ etc.)\footnote{We
call such a $\gs$-algebra a \emph{$\gs$-algebra of sets} - although this
notation does not describe the situation completely.} and let $f : M \to
\RR$ be an $\fram$-measurable function, where we always assume 
that the measurable structure of $\RR$ is given by the $\gs$-algebra
of Borel sets. Then
\[
    \all \ \lir : \ \ef_{\gl} := \urb{f}(]- \∞, \gl]) 
\] 
defines a spectral family $\ef : \RR \to \fram$.\\
~\\
Conversely, if a spectral family $E : \RR \to \fram$ is given, we
obtain a function $f_{E} : M \to \RR$, defined by
\[
    \all \ x \in M : \ f_{E}(x) := \inf \{ \lir \mid x \in \el \}.  
\]
Note that we don't need here any boundedness conditions because the
lattice operations coincide with the usual set operations.   

\begin{proposition}\label{co2}
Let $E : \RR \to \fram$ be a spectral family in the 
sub-$\gs$-algebra \mbox{$\fram \tm pot(M)$.} Then
\begin{displaymath}
    \forall \ \gl \in \RR : \ \overset{-1}{f_{E}}(]-\infty, 
    \gl]) = \el.
\end{displaymath}
\end{proposition}
\emph{Proof:} If $f_{E}(x) ≤ \gl$, then $x \in \emm$ for all $\mu > 
\gl$, hence $x \in \el$. This shows $E^{f_{E}}_{\gl} \tm \el$, and the
converse inclusion is obvious from the definitions. \ \ $\Box$ 

\begin{corollary}\label{co3}
    Let $\fram$ be as above and $E$ a spectral family in $\fram$. 
    Then the function $f_{E} : M \to \RR$ is $\fram$-measurable. 
\end{corollary}

\begin{proposition}\label{co4}
    Let $\fram \tm pot(M)$ be a sub-$\gs$-algebra. Then the spectral 
    families $\RR \to \fram$ are in bijective correspondence to the 
    $\fram$-measurable functions $M \to \RR$:
    \begin{displaymath}
	E^{f_{E}} = E \quad \text{and} \quad  f_{\ef} = f.
    \end{displaymath}
\end{proposition}
\emph{Proof:} $f_{\ef}(x) = \inf \{ \lir \mid x \in \urb{f}(]- \∞,
\gl]) \}$, hence 
\[
     f_{\ef}(x) = \inf \{ \lir \mid f(x) ≤ \gl \} = f(x)
\]     
for all $x \in M$. \ \ $\Box$ \\ 
~\\
This simple result leads immediately to the following

\begin{definition}\label{co5}
    Let $\frA$ be an arbitrary $\gs$-algebra. A spectral family $E$ in
    $\frA$ is called a real valued generalized $\frA$-measurable
    function. 
\end{definition}
~\\
Let $\frA$ be an arbitrary $\gs$-algebra and let $E =
(\el)_{\lir}$ be a \emph{bounded} spectral family in $\frA$. In analogy to the
notion of observable functions in the previous part (\cite{deg4}) we
define

\begin{definition}\label{co6}
    Let $\frA$ be a $\gs$-algebra and $\qA$ the Stone spectrum of $\frA$.
    Then the function $f_{E} : \qA \to \RR$, defined by
    \[
        f_{E}(\frb) := \inf \{ \lir \mid \el \in \frb \},
    \]
    is called the function associated to the bounded spectral family
    $E$.
\end{definition}
The spectral family $E$ is uniquely determined by its associated
function. This was in principle proved for spectral families in a
($\gs$)-complete orthomodular lattice already in part II
(\cite{deg4}), but we like to give here a more direct proof.

\begin{proposition}\label{co7}
    The mapping $E \tto f_{E}$ from the set $\kS_{b}(\LL)$ of all
    bounded spectral families in the $\gs$-complete lattice $\LL$ to
    the set of functions $\ql \to \RR$ is injective.
\end{proposition}
\emph{Proof:} Let $E, F$ be bounded spectral families in $\LL$ such 
that $f_{E} = f_{F}$. Assume that there is some $\gl \irr$ such that
$\el \we (\el \we F_{\gl})^\pp \ne 0$. If $\frb$ is any quasipoint in 
$\LL$ that contains $\el \we (\el \we F_{\gl})^\pp$, then also $\el
\in \frb$ and, therefore, $f_{E}(\frb) ≤ \gl$. But $f_{E}(\frb) < \gl$
would imply $f_{F}(\frb) < \gl$, hence $F_{\gl} \in \frb$, a
contradiction. Thus $f_{F}(\frb) = f_{E}(\frb) = \gl$ for all $\frb
\in \kQ_{\el \we (\el \we F_{\gl})^\pp}(\LL)$, so
\[
    \all \ \mu > \gl : \ F_{\mu} \in \bigcap \kQ_{\el \we (\el \we
    F_{\gl})^\pp}(\LL).
\]  
But $\bigcap \kQ_{\el \we (\el \we F_{\gl})^\pp}(\LL) = 
H_{\el \we (\el \we F_{\gl})^\pp}$, the principal dual ideal generated
by $\el \we (\el \we F_{\gl})^\pp$, so $F_{\mu} ≥ \el \we (\el \we
F_{\gl})^\pp$ for all $\mu > \gl$. Hence also $F_{\gl} ≥ \el \we (\el \we
F_{\gl})^\pp$, a contradiction again. This shows $\el = \el \we
F_{\gl}$, i.e. $\el ≤ F_{\gl}$, for all $\lir$. Since the argument is 
symmetric in $E$ and $F$, we have $E = F$. \ \ $\Box$ \\ 
~\\
Moreover, it is surprisingly easy to see that the function associated 
to a bounded spectral family in a $\gs$-complete orthomodular lattice 
is continuous:

\begin{proposition}\label{co8}
    Let $E$ be a bounded spectral family in a $\gs$-complete orthomodular
    lattice $\LL$. Then the function $f_{E} : \ql  \to \RR$ associated
    to $E$ is continuous.
\end{proposition}
\emph{Proof:} If $f_{E}(\frb_{0}) = \gl$ and $\eps > 0$, take
$\frb \in \kQ_{E_{\gl + \eps}}(\LL) \smm \kQ_{E_{\gl - \eps}}(\LL)$.
Then
\[
    \gl - \eps ≤ f_{E}(\frb) ≤ \gl + \eps, 
\] 
and since $\kQ_{E_{\gl + \eps}}(\LL) \smm \kQ_{E_{\gl - \eps}}(\LL)$
is an open neighbourhood of $\frb_{0}$, we see that  $f_{E}$ is
continuous. \ \ $\Box$ \\
~\\
According to the theorem of Loomis and Sikorski (\cite{Sik}), an
arbitrary $\frA$ is $\gs$-isomorphic to $\frA(M) / \kI$, where $\frA(M)$ 
is a $\gs$-algebra of subsets of a set $M$ and $\kI$ is a suitable $\gs$-ideal in
$\frA(M)$. We have proved in part I (\cite{deg3}, theorem 3.3) that
$\qA$ is homeomorphic to the Gelfand spectrum of the abelian
$C^\ast$-algebra $\kf_{\frA(M)}(M, \CC) / \kf(\kI)$, where 
$\kf_{\frA(M)}(M, \CC)$ is the $C^\ast$-algebra of all bounded
$\frA(M)$-measurable functions $M \to \CC$ and $\kf(\kI)$ is the
norm-closed ideal of those $f \in \kf_{\frA(M)}(M, \CC)$ that vanish
outside some $A \in \kI$. We want to determine the Gelfand
transformation on the selfadjoint part of $\kf_{\frA(M)}(M, \CC) / \kf(\kI)$
as an isomorphism onto $C(\qA, \RR)$. We start with the simpler
situation $\kI = 0$, i.e. $\frA = \fram$. 
        
\begin{proposition}\label{co9}
    Let $\fram$ be a $\gs$-algebra of subsets of a nonempty set $M$ and
    let $\kf_{\fram}(M, \CC)$ be the $C^\ast$-algebra of all bounded
    $\fram$-measurable functions $\gf : M \to \CC$. Then the Gelfand
    spectrum $\gO(\kf_{\fram}(M, \CC))$ of $\kf_{\fram}(M, \CC)$ is
    homeomorphic to the Stone spectrum $\qAm$ of $\fram$ and the
    restriction of the Gelfand transformation to $\kf_{\fram}(M, \RR)$ is
    given, up to the homeomorphism $\qAm \cong \gO(\kf_{\fram}(M, \CC))$, 
    by $\gf \tto f_{\gf}$, where 
    \[
	f_{\gf}(\frb) = \inf \{ \lir \mid \urb{\gf}(]-\∞, \gl]) \in \frb
	\}  
    \]
    for all $\frb \in \qAm$.
\end{proposition}
The \emph{proof} of this proposition is only a slight modification of the
corresponding proof for abelian von Neumann algebras (\cite{deg4}) and so
we can omit the details. \ \ $\Box$ \\
~\\
Let $\kf_{\RR}(\kI)$ be the set of all real valued elements of
$\kf(\kI)$:
\[
    \kf_{\RR}(\kI) = \kf_{\frA(M)}(M, \RR) \cap \kf(\kI). 
\]
If $\gf \in \kf_{\frA(M)}(M, \RR)$ and $\psi \in \kf_{\frA(M)}(M,
\RR)$, then $\gf - \psi \in \kf(\kI)$ if and only if $\gf - \Re \psi \in
\kf_{\RR}(\kI)$ and $\Im \psi \in \kf(\kI)$. Hence the selfadjoint
part of $\kf_{\frA(M)}(M, \CC) / \kf(\kI)$ is isomorphic to
$\kf_{\frA(M)}(M, \RR) / \kf_{\RR}(\kI)$. In what follows, we
determine the Gelfand transformation on $\kf_{\frA(M)}(M, \RR)
/ \kf_{\RR}(\kI)$. \\
~\\
$[\gf] \in \kf_{\frA(M)}(M, \RR) / \kf_{\RR}(\kI)$ defines a spectral family
$E^{[\gf]}$ in $\frA$ in the following way. $\gf \in
[\gf]$ determines\footnote{We do not make a notational distinction between
equivalence classes modulo $\kf_{\RR}(\kI)$ and equivalence classes
modulo $\kf(\kI)$ of real valued functions. Typically, all functions
that occur in the following discussion, will be real valued.} a spectral
family $E^\gf$, given by 
\[
    E^\gf_{\gl} := \urb{\gf}(]- \∞, \gl])
\]
for all $\lir$. If $\psi \in [\gf]$ is another representative, the
definition of equivalence modulo $\kf(\kI)$ implies
\[
    \all \ \lir : \ \urb{\gf}(]- \∞, \gl]) \ \Delta \ \urb{\psi}(]- \∞, \gl])
    \in \kI,
\]
i.e. $E^\gf_{\gl}$ and $E^\psi_{\gl}$ define the same equivalence class 
$[E^\gf_{\gl}]$ modulo $\kI$. Hence 
\[
    E^{[\gf]}_{\gl} := [E^{\gf}_{\gl}]
\] 
is well defined and it is easy to see that $E^{[\gf]} :=
(E^{[\gf]}_{\gl})_{\lir}$ is a spectral family in $\frA$.\\
~\\
We know from corollary 3.3 in \cite{deg3} that $\qA$ is homeomorphic
to $\kQ^\kI (\fram) := \{ \frb \in \qAm \mid \kI^\pp \tm \frb \}$.
So we will regard $\qA$ as this (compact) subset of $\qAm$.
   
\begin{lemma}\label{co10}
    $\bigcap \{ \frb \in \qAm \mid \kI^\pp \tm \frb \} = \kI^\pp$.
\end{lemma}
\emph{Proof:} Assume that there is some $A \in (\bigcap \kQ^\kI (\fram))
\smm \kI^\pp$. If $C \in \kI^\pp$
such that $C \cap A' = \emptyset$, then $C = (C \cap A') \cup (C \cap 
A) = C \cap A$. Hence $C \tm A$, thus $A \in \kI^\pp$, since $\kI^\pp$
is a dual ideal. This contradicts our assumption about $A$. Therefore,
$C \cap A' \ne \emptyset$ for all $C \in \kI^\pp$, hence $\{A'\} \cup 
\kI^\pp$ is contained in a quasipoint $\frb^\sim$. But, by
construction, also $A \in \frb^\sim$, a contradiction again. \ \
$\Box$\\
~\\
\begin{lemma}\label{co11}
    Let $(A_{n})_{\nin}$ be a sequence in $\kI^\pp$. Then also
    $\bigcap_{\nin}A_{n}$ belongs to $\kI^\pp$.  
\end{lemma}
\emph{Proof:} This follows directly from the fact that $\kI$ is a
$\gs$-ideal. \ \ $\Box$ \\

\begin{proposition}\label{co12}
    Let $\gf \in \kf_{\frA(M)}(M, \RR)$, $\egf$ the spectral family
    corresponding to $\gf$ and $f_{\egf} : \qAm \to \RR$ the function
    associated to $\egf$. Then $f_{\egf}$ vanishes on $\qA$ if and
    only if $\gf \in \kf(\kI)$.   
\end{proposition}
\emph{Proof:} If $\gf \in \kf(\kI)$ and $\frb \in \qA$, then $P(\gf) \in
\kI$ and, therefore, $\egf_{0} \in \frb$ but $\egfl \notin \frb$ for
all $\gl < 0$. Hence $f_{\egf}(\frb) = 0$. \\
Conversely, assume that $f_{\egf}$ vanishes on $\qA$. Since $\gf \tto 
f_{\egf}$ is the Gelfand transformation of $\kf_{\frA(M)}(M, \CC)$,
also $f_{E^{\gf_{abs}}}$\footnote{$\gf_{abs}$ denotes the absolute
value of $\gf$.} vanishes on $\qA$. Because of $P(\gf)
= P(\gf_{abs})$, we have $\gf \in \kf(\kI)$ if and only if $\gf_{abs} \in
\kf(\kI)$, hence we may assume that $\gf ≥ 0$, i.e. $\egfl =
\emptyset$ for all $\gl < 0$. Since $f_{\egf}$ vanishes on $\qA$, 
$\egfl \in \frb$ for all $\frb \in \qA$ and all $\gl > 0$. This implies $\egfl
\in \kI^\pp$ for all $\gl > 0$ by lemma \ref{co10}, hence $\egf_{0} \in
\kI^\pp$ by lemma \ref{co11}. But then $\urb{\gf}(0) \in \kI^\pp$, i.e. 
$\gf \in \kf(\kI)$. \ \ $\Box$ \\    
~\\
Consider the mapping
\[
    \begin{array}{cccc}
	 & \kf_{\frA(M)}(M, \RR) & \to & C(\qA, \RR)  \\
	 & \gf & \tto & f_{\egf}\mid_{\qA}.
    \end{array}
\]
As in the case of abelian von Neumann algebras, we can complexify this
mapping to obtain a mapping
\[
    G : \kf_{\frA(M)}(M, \CC) \to C(\qA)
\]
which is the Gelfand transformation of $\kf_{\frA(M)}(M, \CC)$
followed by the restriction to $\qA$. Proposition \ref{co12} implies that
$\kf(\kI)$ is the kernel of $G$. Hence $G$ induces an injective
$\ast$-homomorphism
\[
    \gG : \kf_{\frA(M)}(M, \CC) / \kf(\kI) \to C(\qA)
\] 
of abelian $C^\ast$-algebras. We prove next that $\gG$ is surjective, 
too. Let $\psi \in C(\qA, \RR)$. Since $\qA$ is a closed subset of $\qAm$,
Tietze's extension theorem implies that $\psi$ can be extended to a
function $\psi_{1} \in C(\qAm, \RR)$. $\psi_{1}$ has the form $\psi_{1} =
f_{E^{\gf_{1}}}$ for some $\gf_{1} \in \kf_{\frA(M)}(M, \RR)$. If
$\psi_{1}, \psi_{2} \in C(\qAm, \RR)$ are two such extensions of $\psi$,
and if $\gf_{1}, \ \gf_{2} \in \kf_{\frA(M)}(M, \RR)$ are chosen so that
$\psi_{1} = f_{E^{\gf_{1}}}$ and $\psi_{2} = f_{E^{\gf_{2}}}$, then
$\gf_{1} - \gf_{2} \in \kf(\kI)$. Hence there is a unique $[\gf] \in 
\kf_{\frA(M)}(M, \RR) / \kf_{\RR}(\kI)$ such that $\psi = \gG([\gf]) =
f_{\egf}\mid_{\qA}$. The complex case is a direct consequence.\\
~\\
As we showed above, $[\gf] \in \kf_{\frA(M)}(M, \RR) / \kf_{\RR}(\kI)$
defines a spectral family $E^{[\gf]}$ in $\frA$ by
\[
    \all \ \lir : \ E^{[\gf]}_{\gl} := [\egfl].
\]

\begin{lemma}\label{co13}
    If $[\gf] \in \kf_{\frA(M)}(M, \RR) / \kf_{\RR}(\kI)$, then 
    \[
	f_{E^{[\gf]}} = f_{\egf}\mid_{\qA},
    \]
    where we have identified $\qA$ with $\kQ^{\kI}(\frA(M)) := 
    \{ \frb \in \qAm \mid \kI^\pp \tm \frb \}$.
\end{lemma}
\emph{Proof:} This follows from the following fact, already used in
the proof of corollary 3.3 in \cite{deg3}: If $[A] \in \frA$ and if $B
\in \frA(M)$ is any representative of $[A]$, then $[A] \in \frb$
(considered as a quasipoint in $\frA$) if and only if $B \in \frb$
(considered as an element of $\kQ^\kI(\frA(M))$). \ \ $\Box$ \\ 
~\\
The foregoing results show that we have proved the following

\begin{theorem}\label{co14}
    Let $\frA$ be a $\gs$-algebra, represented as a quotient $\fram / 
    \kI$ of a $\gs$-algebra $\fram$ of subsets of a set $M$ modulo a
    $\gs$-ideal $\kI$ in $\fram$. Then the Gelfand transformation of
    the $C^\ast$-algebra $\kf_{\frA(M)}(M, \CC) / \kf(\kI)$,
    restricted to the selfadjoint part $\kf_{\frA(M)}(M, \RR) /
    \kf_{\RR}(\kI)$, is given by
    \[
	\begin{array}{cccc}
	    \gG : & \kf_{\frA(M)}(M, \RR) / \kf_{\RR}(\kI) & \to &
	    C(\qA, \RR)  \\
	     & [\gf] & \tto & f_{E^{[\gf]}}. 
	\end{array}
    \] 
    Moreover, if $\qA$ is identified with $\kQ^\kI(\frA(M))$, we can
    write
    \[
	f_{E^{[\gf]}} = f_{\egf}\mid_{\qA},
    \]
    where $\gf \tto f_{\egf}$ is the Gelfand transformation on
    $\kf_{\frA(M)}(M, \RR)$.
\end{theorem}
We have already mentioned that for an abstract $\gs$-algebra $\frA$,
spectral families in $\frA$ are the adequate substitutes for
$\frA$-measurable functions. If $\frA$ is represented as a quotient
$\fram / \kI$ with $\fram$ and $\kI$ as above, we shall show that each
\emph{bounded} spectral family in $\frA$ is the quotient modulo $\kI$
of a spectral family in $\fram$. This means that every bounded
spectral family in $\frA$ can be lifted to the spectral family of a
bounded $\fram$-measurable function. The proof rests on the foregoing
theorem.

\begin{corollary}\label{co15}
    Let $E$ be a bounded spectral family in a $\gs$-algebra $\frA$ and let
    $\frA = \fram / \kI$, where $\fram$ is a $\gs$-algebra of subsets 
    of some set $M$ and $\kI$ is a $\gs$-ideal in $\fram$. Then there 
    is some $\gf \in \kf_{\frA(M)}(M, \RR)$ such that
    \[
	\all \ \lir : \ \el = [\egfl].
    \]
    $\gf$ is unique up to equivalence modulo $\kf_{\RR}(\kI)$. 
\end{corollary}
\emph{Proof:} Let $f_{E} : \qA \to \RR$ be the function associated to 
$E$. $f_{E}$ is continuous, so it is the Gelfand transform of a unique
$[\gf] \in \kf_{\frA(M)}(M, \RR) / \kf_{\RR}(\kI)$, i.e.
\[
    f_{E} = f_{E^{[\gf]}}.
\]
According to propositions \ref{co7}, \ref{co8}, the mapping $E \tto f_{E}$
from the set of bounded spectral families in $\frA$ to $C(\qA, \RR)$ is
injective, hence $E = E^{[\gf]}$.\ \ $\Box$ \\ 
~\\
We finish this section with an extension of the previous results to
\emph{complex valued} (generalized) measurable functions. \\
~\\
Let $\LL$ be a ($\gs$-) complete lattice and let $E =
(\el)_{\lir}$ be a bounded spectral family in $\LL$. We have defined
the function $f_{E} : \ql \to \RR$ by $f_{E}(\frb) := \inf \{ \lir
\mid \el \in \frb \}$, and we want to generalize this concept to
\emph{$n$-parameter spectral families}. \\
To motivate the requirements of the general definition, we consider
first an important special case. Let $\LL =\pr$ for a finite von Neumann
algebra $\rr$ and let $E^1, \ldots, E^n$ be spectral families in $\pr$.
Then the mapping 
\[
    \begin{array}{cccc}
	G : & \RR^n & \to & \pr  \\
	 & (\gl_{1}, \ldots, \gl_{n}) & \tto & E^1_{\gl_{1}} \we \cdots
	 \we E^n_{\gl_{n}},
    \end{array}
\]
has the following properties:
\begin{enumerate}
    \item  [(i)] $G(\gl_{1}, \ldots, \gl_{n}) \we G(\mu_{1}, \ldots,
    \mu_{n}) = G(\nu_{1}, \ldots, \nu_{n})$, where $\nu_{k} = \min \{ 
    \gl_{k}, \mu_{k} \}$ for $k = 1, \ldots, n$.  

    \item  [(ii)] $\We_{\gl_{1} < \mu_{1}, \ldots, \gl_{n} < \mu_{n}}
    G(\mu_{1}, \ldots, \mu_{n}) = G(\gl_{1}, \ldots, \gl_{n})$ for all
    $\gl_{1}, \ldots, \gl_{n} \irr$.

    \item  [(iii)] $\We_{\lir}G(\gl_{1}, \ldots, \gl_{k - 1}, \gl,
    \gl_{k + 1}, \ldots, \gl_{n}) = 0$ for all $k = 1, \ldots, n$, and
    $\Ve_{\gl_{1}, \ldots, \gl_{n} \irr}G(\gl_{1}, \ldots, \gl_{n}) =
    I$.
\end{enumerate}
Note that only the last property requires the finiteness of $\rr$ if
we consider \emph{arbitrary}, not necessarily bounded, spectral families.

\begin{definition}\label{co16}(\cite{na})\\
    Let $\LL$ be a complete (orthomodular) lattice. An $n$-parameter
    spectral family is a mapping
    \[
	\begin{array}{cccc}
	    E : & \RR^n & \to & \LL  \\
	     & (\gl_{1}, \ldots, \gl_{n}) & \tto & E_{\gl_{1}, \ldots,
	     \gl_{n}}
	\end{array}
    \]
    with the following properties:
    \begin{enumerate}
	\item  [(i)] $E_{\gl_{1}, \ldots, \gl_{n}} \we E_{\mu_{1}, \ldots,
	    \mu_{n}} = E_{\nu_{1}, \ldots, \nu_{n}}$, where $\nu_{k} =
	\min \{ \gl_{k}, \mu_{k} \}$ for $k = 1, \ldots, n$.  

	\item  [(ii)] $\We_{\gl_{1} < \mu_{1}, \ldots, \gl_{n} < \mu_{n}}
	    E_{\mu_{1}, \ldots, \mu_{n}} = E_{\gl_{1}, \ldots,
	    \gl_{n}}$ for all $\gl_{1}, \ldots, \gl_{n} \irr$.
	
	\item  [(iii)] $\We_{\lir}E_{\gl_{1}, \ldots, \gl_{k - 1}, \gl,
	    \gl_{k + 1}, \ldots, \gl_{n}} = 0$ for all $k = 1, \ldots,
	    n$,\\ and
	   $\Ve_{\gl_{1}, \ldots, \gl_{n} \irr}E_{\gl_{1}, \ldots,
	   \gl_{n}} = I$.
    \end{enumerate}
    $E$ is called bounded if there are $m, M \irr$ such that $E_{\gl_{1},
    \ldots, \gl_{n}} = 0$, if $\gl_{k} ≤ m$ for at least one $k = 1,
    \ldots, n$, and $E_{\gl_{1}, \ldots, \gl_{n}} = I$ if $\gl_{1}, \ldots,
    \gl_{n} ≥ M$.
\end{definition}
In what follows, we restrict our discussion to $2$-parameter spectral 
families. This is no real restriction, for the general case requires
only more typing effort. For apparent reasons, $2$-parameter spectral 
families are called \emph{complex spectral families.}\\
~\\
Let $E = (E_{\gl_{1}, \gl_{2}})_{\gl_{1}, \gl_{2} \irr}$ be a bounded 
complex spectral family. We want to define the function associated to 
$E$ as a function 
\[
     f_{E} : \ql \to \CC,
\] 
where we identify $\CC$ with $\RR^2$. If $\frb \in \ql$, define
\[
    f_{E, 1}(\frb) := \inf \{ \lir \mid \ex \ \mu \irr : \ E_{\gl, \mu}
    \in \frb \}
\]
and, symmetrically,
\[
    f_{E, 2}(\frb) := \inf \{ \mu \irr \mid \ex \ \gl \irr : \ E_{\gl, \mu}
    \in \frb \}.
\]
Eventually, we define
\[
    f_{E}(\frb) := f_{E, 1}(\frb) + i f_{E, 2}(\frb).
\]

\begin{proposition}\label{co17}
    Let $E^1, E^2$ be bounded spectral families in $\LL$ and let 
    \[
	\all \ \gl, \mu \irr : \ E_{\gl, \mu} := E^1_{\gl} \we
	E^2_{\mu}.
    \]
    Then $E := (E_{\gl, \mu})_{\gl, \mu \irr}$ is a bounded complex
    spectral family, and the function associated to $E$ is
    \[
	f_{E} = f_{E^1} + i f_{E^2}.
    \]
\end{proposition}
\emph{Proof:} Let $M_{k} \irr$ such that $E^k_{M_{k}} = I, \ (k = 1,
2)$. Then $E_{\gl, M_{2}} = E^1_{\gl}$ and $E_{M_{1}, \mu} =
E^2_{\mu}$ for all $\gl, \mu \irr$. Since $E_{\gl, \mu} \in \frb$ if
and only if $E^1_{\gl}, E^2_{\mu} \in \frb$, and since the mapping
$(\gl, \mu) \tto E_{\gl, \mu}$ is increasing in both variables, we
obtain
\[
    f_{E, 1}(\frb) = \inf \{ \lir \mid E_{\gl, M_{2}} \in \frb \} =
    f_{E^1}(\frb), 
\]
and similarly $f_{E, 2} = f_{E^2}$. \ \ $\Box$ \\
~\\

An important special case is $\LL = \pr$ for a von Neumann algebra
$\rr$. If $A \in \rr, \ A = A_{1} + i A_{2}$ its decomposition
into selfadjoint parts, and if $E^k$ is the spectral resolution of
$A_{k} \ \ (k = 1, 2)$, then $A$ is represented by the complex
spectral family $E : (\gl, \mu) \tto E_{\gl, \mu} := E^1_{\gl} \we
E^2_{\mu}$. If $A$ is normal, then $E_{\gl, \mu} = E^1_{\gl} 
E^2_{\mu}$ and $E$ is the \emph{complex} spectral resolution of $A$ :
\[
A = \int_{\CC}(\gl + i \mu)dE_{\gl, \mu}.
\]
We know that $sp(A) = im f_{A}$ for selfadjoint $A$. Note,
however, that this is not true, in general, for non-selfadjoint
$A$. This can be seen already in the following very simple
example:\\
Let $A := \begin{pmatrix}
1  &  0  \\
0  &  i
\end{pmatrix}
= P_{\CC e_{1}} + i P_{\CC e_{2}} \in \kL(\CC^2)$. Then $sp(A) =
\{1, i\}$, but for all lines $\CC x \notin \{\CC e_{1}, \CC
e_{2}\}$ we have $f_{P_{\CC e_{1}}}(\frb_{\CC x}) = f_{P_{\CC
e_{2}}}(\frb_{\CC x}) = 1$. Hence $f_{A}(\frb_{\CC x}) = 1 + i
\notin sp(A)$. If we consider $A$ as an element of the maximal
abelian subalgebra $\mm$ generated by $P_{\CC e_{1}}$, then
$f_{A}(\qmm) = sp(A)$, since $f_{A}\mid_{\qmm}$ is the Gelfand
transform of $A$. Hence, contrary to the selfadjoint case,  the
image of $\dr$ by $f_{A}$ may be strictly smaller than
$f_{A}(\qr)$. \\
 
\begin{definition}\label{co18}
    A complex spectral family $E = (E_{\gl, \mu})_{\gl, \mu}$ in a
    complete lattice $\LL$ is called decomposable, if there are spectral
    families $E^1, E^2$ in $\LL$ such that
    \[
	\all \ \gl, \mu \irr : \  E_{\gl, \mu} = E^1_{\gl} \we
	E^2_{\mu}.
    \]
\end{definition}
Let $E = (E_{\gl, \mu})_{\gl, \mu \irr}$ be a bounded decomposable
complex spectral family. Then the spectral families $E^1, E^2$ are
bounded and uniquely determined by $E$. Moreover, there are $b_{1},
b_{2} \irr$ such that $E^1_{\gl} = I = E^2_{\mu}$ for all $\gl ≥
b_{1}, \mu ≥ b_{2}$, hence $E_{\gl, \mu} = E^2_{\mu}$ for all $\gl ≥
b_{1}$ and $E_{\gl, \mu} = E^1_{\gl}$ for all $\mu ≥ b_{2}$. This
property can be formulated in the following way:
\[
    \ex \ b_{1} \irr \ \ \all \ \mir : \ \gl \tto E_{\gl, \mu} \quad
    \text{is constant on} \quad [b_{1}, \∞ [, 
\]
and 
\[
    \ex \ b_{2} \irr \ \ \all \ \lir : \ \mu \tto E_{\gl, \mu} \quad
    \text{is constant on} \quad [b_{2}, \∞ [.
\]

\begin{lemma}\label{co19}
    Let $E = (E_{\gl, \mu})_{\gl, \mu \irr}$ be a bounded complex 
    spectral family. Then $E$ is decomposable if and only if it
    satisfies the following two properties:
    \begin{enumerate}
	\item  [(i)] $\ex \ b_{1} \irr \ \ \all \ \mir : \ \gl \tto E_{\gl, \mu}
	\quad \text{is constant on} \quad [b_{1}, \∞ [$,
    
	\item  [(ii)] $\ex \ b_{2} \irr \ \ \all \ \lir : \ \mu \tto E_{\gl, \mu}
	\quad \text{is constant on} \quad [b_{2}, \∞ [.$
    \end{enumerate} 
\end{lemma}
\emph{Proof:} If $E$ is decomposable, choose $b_{k} \irr$ such that
$E^k_{b_{k}} = I \ \ (k = 1, 2)$. \\
Conversely, define for $\gl, \mu \irr$
\[
    E^1_{\gl} := E_{\gl, b_{2}} \quad \text{and} \quad E^2_{\mu} := 
    E_{b_{1}, \mu}.
\]
Then
\[
    E_{\gl, \mu} = E^1_{\gl} \we E^2_{\mu}
\]
for all $\gl, \mu \irr$: if $\gl ≤ b_{1}, \mu ≤ b_{2}$, this follows
directly from the definition of complex spectral families. If $\gl >
b_{1}$, then $E_{\gl, \mu} = E^2_{\mu}$ and $E^1_{\gl} = I$ by $(i),
(ii)$. The argument for $\mu > b_{2}$ is similar. \ \ $\Box$ \\

\begin{proposition}\label{co20}
    Every bounded complex spectral family $E$ in a complete lattice
    $\LL$ is decomposable.
\end{proposition}
\emph{Proof:} Let $b \irr$ be an upper bound for $E$, i.e. $E_{\gl, \mu}
= I$ for all $\gl, \mu ≥ b$. If $\gl ≥ b$ and $\mu \irr$, then 
\[
    E_{\gl, \mu} = E_{\gl, \mu} \we E_{b, b} = E_{b, \min\{b, \mu\}},
\]
hence $\gl \tto E_{\gl, \mu}$ is constant on $[b, \∞ [$. Analogously, 
$\mu \tto E_{\gl, \mu}$ is constant on $[b, \∞[$ for all $\lir$. \ \
$\Box$ \\  
~\\
Note that the foregoing result rests on the strong monotonicity
condition $(i)$ in definition \ref{co16}. If we would impose the
weaker, but also completely natural condition
\[
    E_{\gl, \mu} ≤ E_{\gl', \mu'} \quad \text{for} \quad \gl ≤ \gl',
    \mu ≤ \mu',
\]
proposition \ref{co20} were not true. \\

\begin{corollary}\label{co21}
    Let $E$ be a bounded complex spectral family in a complete
    orthomodular lattice $\LL$. Then the function $f_{E} : \ql \to
    \CC$, associated to $E$, is continuous.
\end{corollary}
\emph{Proof:} $E$ is decomposable, hence, according to proposition
\ref{co17}, $f_{E} = f_{E^1} + i f_{E^2}$ with bounded spectral families
$E^1, E^2$ in $\LL$ such that $E_{\gl, \mu} = E^1_{\gl} \we E^2_{\mu}$
for all $\gl, \mu \irr$. By proposition \ref{co8}, the functions
$f_{E^1}$ and $f_{E^2}$ are continuous. \ \ $\Box$

\begin{corollary}\label{co22}
    The decomposition of a bounded complex spectral family $E$ in a
    complete orthomodular lattice $\LL$ into spectral families $E^1,
    E^2$ is unique.
\end{corollary}
\emph{Proof:} The decomposition $f_{E} = f_{E^1} + i f_{E^2}$ of the
function $f_{E}$ associated to $E$ into real and imaginary part is
unique and, by proposition \ref{co7}, $f_{E^k}$ determines $E^k$. \ \
$\Box$ \\
~\\
As an application of the foregoing results we obtain

\begin{theorem}\label{co23}
    The set $\kS_{b}(\CC, \frA)$ of all bounded complex spectral families
    in a $\gs$-algebra $\frA$ forms an abelian $C^\ast$-algebra which
    is $\ast$-isomorphic to $C(\qA)$. 
\end{theorem}
\emph{Proof:} We use the bijection
\[
\begin{array}{cccc}
    f_{\ast} : & \kS_{b}(\CC, \frA) & \to & C(\qA)  \\
     & E & \tto & f_{E}
\end{array}
\]
to define the $C^\ast$-algebraic structure of $\kS_{b}(\CC, \frA)$. If $E, F
\in \kS_{b}(\CC, \frA)$ and $\ga \in \CC$, define
\begin{eqnarray*}
    E + F & := & f_{\ast}^{-1}(f_{E} + f_{F}),  \\
    EF & := & f_{\ast}^{-1}(f_{E}f_{F}),  \\
    \ga E & := & f_{\ast}^{-1}(\ga f_{E}),  \\
    E^\ast & := & f_{\ast}^{-1}(\overline{f_{E}}),  \\
    |E| & := & |f_{E}|_{\∞}.
\end{eqnarray*}
 Then $f_{\ast}$ is a $\ast$-isomorphism by definition. \ \ $\Box$ 
~\\
~\\
We finish this section by showing up a {\bf spectral theorem for
measurable functions.}\\
~\\
Let $M$ be a nonvoid set, $\fram \tm pot(M)$ a $\gs$-algebra and $\gf 
: M \to \RR$ a $\fram$-measurable function with spectral family
$E^{\gf}$. Since 
\[
    \all \ \lir : \ \egfl = \urb{\gf}(]- \∞, \gl]),
\]
we obtain for all $\gl < \mu$:
\[
    \all \ x \in \egf_{\mu} \we (\egfl)^{\pp} : \ \gl < \gf(x) ≤ \mu.
\]
Let $\eps > 0$ and let $(\gl_{k})_{k \in \ZZ}$ be a sequence of
real numbers such that $\gl_{k + 1} - \gl_{k} = \eps$ for $k \in \ZZ$.
By construction, the step function 
\[
    s := \sum_{k \in \ZZ}\gl_{k}(\chi_{\egf_{\gl_{k + 1}}} -
    \chi_{\egf_{\gl_{k}}})
\]
satisfies
\[
    |\gf - s|_{\∞} ≤ \eps.
\]
Since the mapping 
\[
    \begin{array}{cccc}
	\chi_{\egf} : & \RR & \to & \kf_{\fram}(M, \RR)  \\
	 & \gl & \tto & \chi_{\egf_{\gl}} 
    \end{array}
\]
is increasing and $\gf$ can be uniformly approximated by step functions 
of the form above, we can write $\gf$ as a \emph{Riemann-Stieltjes}
integral:
\[
    \gf = \int_{\RR}\gl d\chi_{\egfl}.
\]
This is quite analogous to the spectral representation of bounded
selfadjoint operators of a Hilbert space. We have therefore proved:

\begin{proposition}\label{co23a}
    ({\bf Spectral theorem for measurable functions})\\
    Let $\gf : M \to \RR$ be an $\fram$-measurable function with spectral
    family $\egf$. Then 
    \[
	\gf = \int_{\RR}\gl d\chi_{\egfl}
    \]
    in the sense of a Riemann-Stieltjes integral.\\
    Conversely, if $E$ is a spectral family in $\fram$, the
    $\fram$-measurable function $\gf_{E} : M \to \RR$ induced by $E$
    can be represented as the Riemann-Stieltjes integral
    \[
	\gf_{E} = \int_{\RR}\gl d\chi_{E_{\gl}}
    \] 
    with respect to $\chi_{E} : \gl \tto \chi_{\el}$.
\end{proposition}
~\\
Now let $\frA$ be an abstract $\gs$-algebra, $E$ a bounded spectral
family in $\frA$ and $f_{E} : \qA \to \RR$ the continuous function
associated to $E$. Then 
\[
    \begin{array}{cccc}
	\chi^{\kQ}_{E} : & \RR & \to & C(\qA, \RR)  \\
	 & \gl & \tto & \chi_{\kQ_{\el}(\frA)} 
    \end{array}
\]
is an increasing mapping that equals $1$ for $\gl ≥ b$ and $0$ for
$\gl ≤ a$, where $a, b$ are the bounds of the spectral family $E$.
Therefore the same reasoning as above shows that 
\[
    f_{E} = \int_{\RR}\gl d\chi^{\kQ}_{E_{\gl}}.
\]
This is the spectral theorem for bounded generalized $\frA$-measurable
functions.

\section{Continuous Functions}
\label{cob}
\pagestyle{myheadings}
\markboth{Classical Observables}{Continuous Functions}

In this section we study spectral families in $\kT(M)$ that 
correspond to continuous real valued functions on a Hausdorff
topological space $M$. Recall that in the lattice $\kT(M)$ the 
(infinite) lattice operations are given by
\begin{displaymath}
     \bigvee_{\ga \in A}U_{\ga} = \bigcup_{\ga \in A}U_{\ga}
\end{displaymath}
and
\begin{displaymath}
      \bigwedge_{\ga \in A}U_{\ga} = int(\bigcap_{\ga \in A}U_{\ga}),
\end{displaymath}
where $int N$ denotes the interior of the subset $N$ of $M$. It is the
occurence of the interior in the definition of the infinite meet that 
causes some technical difficulties.\\
~\\
We begin with some simple examples:

\begin{example}\label{co24}
    The following settings define spectral families $E^{id}, 
    E^{abs}, E^{ln}, E^{step}$ in $\kT(\RR)$:
    \begin{eqnarray}
	E^{id}_{\gl} & := & ]-\infty, \gl[,
	\label{1}  \\
	 E^{abs}_{\gl} & := & ]-\gl, \gl[
	\label{2}  \\
	E^{ln}_{\gl} & := & ]-\exp(\gl), \exp(\gl)[
	\label{3}  \\
	E^{step}_{\gl} & := & ]-\infty, \lfloor\gl\rfloor[
	\label{4}
    \end{eqnarray}
    where $\lfloor\gl\rfloor$ denotes the ``floor of $\gl \in \RR$'':
    \begin{displaymath}
	\lfloor\gl\rfloor = \max \{n \in \ZZ \mid n \leq \gl \}.
    \end{displaymath}
\end{example}
The names of these spectral families sound somewhat crazy at the 
moment, but we will justify them soon. \\
~\\
In close analogy to the case of spectral families in a $\gs$-algebra
$\fram$ of subsets of $M$, each spectral family in $\kT(M)$ induces a 
function on a subset of $M$.

\begin{definition}\label{co25}
    Let $E : \RR \to \kT(M)$ be a spectral family in $\kT(M)$. Then
    \begin{displaymath}
	\kD(E) := \{x \in M \mid \exists \ \gl \in \RR : \ x \notin 
	 \el) \}
    \end{displaymath}
    is called the {\bf admissible domain of $E$}.
\end{definition}
Note that 
\[
\kD(E) = M \smm \bigcap_{\gl \in \RR} \el)
\]
may be different from $M$ because it is possible that $\bigcap_{\lir}\el
\ne \emptyset$, although $int \bigcap_{\lir}\el = \emptyset$. The 
spectral family $E^{ln}$ is a simple example:
\begin{displaymath}
    \forall \ \gl \in \RR : \ 0 \in E^{ln}_{\gl}.
\end{displaymath}
Clearly, if $E$ is bounded from below, then $\kD(E) = M$.
  
\begin{remark}\label{co26}
    The admissible domain $\kD(E)$ of a spectral family $E : \RR 
    \to \kT(M)$ is dense in $M$.
\end{remark}
\emph{Proof:} $\kD(E) = \emptyset$ means that $\el = M$ for
all $\gl \in \RR$, contradicting $\We_{\gl \in \RR}\el =
\emptyset.$
Moreover $U \cap \kD(E) = \emptyset$ for some $U \in \kT(M)$
implies that $U \tm \bigwedge_{\gl \in \RR}\el = \emptyset$.
This shows that $\kD(E)$ is dense in $M$. \ \ $\Box$ \\
~\\
Each spectral family $E : \RR \to \kT(M)$ induces a function 
$f_{E} : \kD(E) \to \RR$:

\begin{definition}\label{co27}
    Let $E : \RR \to \kT(M)$ be a spectral family with admissible 
    domain $\kD(E)$. Then the function $f_{E} : \kD(E) \to 
    \RR$, defined by
    \begin{displaymath}
	\forall \ x \in \kD(E) : \ f_{E}(x) := \inf \{\gl \in \RR 
	\mid x \in \el \},
    \end{displaymath}
    is called the {\bf function induced by $E$}.
\end{definition}
In complete analogy to the operator case we define the spectrum of 
a spectral family $E : \RR \to \LL$ in a complete lattice $\LL$:

\begin{definition}\label{co28}
    Let $E : \RR \to \LL$ be a spectral family. Then
    \begin{displaymath}
	R(E) := \{\gl \in \RR \mid E \  \text{is constant on a 
	neighborhood of}\  \gl \}
    \end{displaymath} 
    is called the {\bf resolvent set} of $E$, and
    \begin{displaymath}
	sp(E) := \RR \setminus R(E)
    \end{displaymath}
    is called the {\bf spectrum} of $E$.
\end{definition}

Obviously $sp(E)$ is a closed subset of $\RR$.

\begin{proposition}\label{co29}
    Let $f_{E} : \kD(E) \to \RR$ be the function induced by the 
    spectral family $E : \RR \to \kT(M)$. Then
    \begin{displaymath}
	sp(E) = \overline{im f_{E}}.
    \end{displaymath}
\end{proposition}
\emph{Proof:} It is obvious that $R(E) \tm M \smm im f_{E}$
holds. This means that $\overline{im f_{E}}$ is a subset of $sp(E)$. 
\\  If $\gl \notin \overline{im f_{E}}$ then $]\gl - \gd, \gl + \gd [
\ \cap \ im f_{E} = \emptyset$ for some $\gd > 0$. Assume that there are
$\gl_{1}, \gl_{2} \in \RR$ such that $\gl - \gd < \gl_{1} < \gl_{2} < 
\gl + \gd$ and $E_{\gl_{1}} \neq E_{\gl_{2}}$. Then $f_{E}(x) \in
[\gl_{1}, \gl_{2}]$ for all $x \in \gs(\gl_{2}) \smm E_{\gl_{1}}$, a 
contradiction. This shows that $sp(E)$ is contained in
$\overline{im f_{E}}$. \ \ $\Box$ \\
~\\
The functions induced by our foregoing examples are
\begin{eqnarray}
    f_{E^{id}}(x) & = & x
    \label{1}  \\
    f_{E^{abs}}(x) & = & |x|
    \label{2}  \\
    f_{E^{ln}}(x) & = & \ln{|x|} \quad  \text{and} \ \kD(E^{ln}) = \RR 
    \setminus\{0\}
    \label{3}  \\
    f_{E^{step}} & = & \sum_{n \in \ZZ}n\chi_{[n, n + 1[}
    \label{4}
\end{eqnarray}
There is a fundamental difference between the spectral families 
$E^{id}, E^{abs}, E^{ln}$ on the one side and $E^{step}$ 
on the other. The function induced by $E^{step}$ is not 
continuous. This fact is mirrored in the spectral families: the 
first three spectral families have the property
\begin{displaymath}
    \forall \ \gl < \mu : \ \overline{\el} \tm \emm.
\end{displaymath}
Obviously $E^{step}$ fails to have this property.

\begin{definition}\label{co30}
    A spectral family $E : \RR \to \kT(M)$ is called {\bf strongly
    regular} if
    \begin{displaymath}
	\forall \ \gl < \mu : \ \overline{\el} \tm \emm
    \end{displaymath}
    holds.
\end{definition}      
Using the pseudocomplement $U^{c} := M \setminus \bar{U} \quad (U 
\in \kT(M))$ we can express the condition of strong regularity in purely 
lattice theoretic terms as
\begin{displaymath}
    \forall \ \gl < \mu : \ \el^{c} \cup \emm = M. 
\end{displaymath}
~\\
If $E = (E_{\gl, \mu})_{\gl, \mu \irr}$ is a \emph{bounded}
complex spectral family in $\ktm$, strong regularity of $E$ can be
defined in two equivalent manners. We know from proposition
\ref{co20} that $E$ is decomposable, i.e. $E_{\gl, \mu} =
E^{1}_{\gl} \cap E^{2}_{\mu}$ for all $\gl, \mu \irr$, where
$E^{1}, E^{2}$ are (necessarily bounded) spectral families in
$\ktm$. Hence, since this decomposition is unique, it is natural to
call $E$ strongly regular if the spectral families $E^{1}, E^{2}$ are
strongly regular. If $E$ is strongly regular in this sense, we
obtain for $\gl < \gl', \mu < \mu'$:
\[
\overline{E_{\gl, \mu}} \tm \overline{E^{1}_{\gl}} \cap
\overline{E^{2}_{\mu}} \tm E^{1}_{\gl'} \cap E^{2}_{\mu'} =
E_{\gl', \mu'}.
\]
Conversely, the condition
\[
\all \ \gl < \gl', \mu < \mu' : \ \overline{E_{\gl, \mu}} \tm 
E_{\gl', \mu'},
\]
implies that $E^{1}$ and $E^{2}$ are strongly regular:
If $b \irr$ is chosen such that $E^{1}_{\gl} = E^{2}_{\mu} =
M$ for all $\gl, \mu ≥ b$, then
\[
\all \ \gl < \gl' : \ \overline{E^{1}_{\gl}} =
\overline{E^{1}_{\gl} \cap E^{2}_{b}} \tm E^{1}_{\gl'}.  
\]
Similarly, $\overline{E^{2}_{\gl}} \tm E^{2}_{\mu}$ for all $\gl <
\mu$. Hence $E^{1}, E^{2}$ are strongly regular spectral families.
Therefore, we can define strong regularity of a bounded complex spectral
family in two equivalent ways.  

\begin{definition}\label{co30a}
    A bounded complex spectral family $E$ in $\ktm$ is called strongly
    regular, if one of the following equivalent properties is
    satisfied:
    \begin{enumerate}
        \item  [(i)] $\all \ \gl < \gl', \mu < \mu' : \ \overline{E_{\gl, \mu}} \tm 
                          E_{\gl', \mu'}$.

        \item  [(ii)] The bounded spectral families $E^{1}, E^{2}$ in 
	the decomposition of $E$ are strongly regular.
    \end{enumerate}
\end{definition}

\begin{remark}\label{co31}
    The admissible domain $\kD(E)$ of a strongly regular  spectral 
    family $E : \RR \to \kT(M)$ is an open (and dense) subset of $M$.
\end{remark}
\emph{Proof:} Let $x \in \kD(E)$ and choose $\gl_{0} \in \RR$ such
that $x \notin E_{\gl_{0}}$. Because $E$ is strongly regular  we have $x
\notin \overline{\el}$ for all $\gl < \gl_{0}$. Let $U$ be an
open neighborhood of $x$ that is contained in the complement of 
$\overline{\el}$. Then $y \notin \el$ for all $y \in U$,
i.e. $U \tm \kD(E)$. \ \ $\Box$ \\
~\\
The name ``strongly regular  spectral family'' is motivated by the following
result:

\begin{remark}\label{co32}
    If $E : \RR \to \kT(M)$ is a strongly regular  spectral family, then, for 
    all $\gl \in \RR$, $\el$ is a regular open set, i.e.
    \begin{displaymath}
	\el^{cc} = \el.
    \end{displaymath}
\end{remark}
\emph{Proof:}
For all $\gl \in \RR$ and all $\mu > \gl$ we have
\[
    \el \tm \overline{\el} \tm \emm
\]
and therefore
\[
    \el \tm int (\overline{\el}) \tm \We_{\mu > \gl}
    \emm = \el. \ \ \Box
\]
~\\
The importance of strongly regular  spectral families becomes manifest 
in the following

\begin{theorem}\label{co33}
    Let $M$ be a toplogical space. Then every continuous function $f : 
    M \to \RR$ induces a strongly regular  spectral family $E^{f} : \RR 
    \to \kT(M)$ by
    \begin{displaymath}
	\forall \ \gl \in \RR : \ E^{f}(\gl) := int(\overset{-1}{f}(]-\infty,
	\gl])).
    \end{displaymath}
    The admissible domain $\kD(E^{f})$ equals $M$ and the function 
    $f_{E^{f}} : M \to \RR$ induced by $E^{f}$ is $f$. 
    Conversely, if $E : \RR \to \kT(M)$ is a strongly regular  spectral 
    family, then the function
    \begin{displaymath}
	f_{E} : \kD(E) \to \RR
    \end{displaymath}
    induced by $E$ is continuous and the induced spectral family 
    $E^{f_{E}}$ in $\kT(\kD(E))$ is the restriction of $E$ to 
    the admissible domain $\kD(E)$:
    \begin{displaymath}
	\forall \gl \in \RR : \ E^{f_{E}}_{\gl} = \el \cap \kD(E). 
    \end{displaymath}
\end{theorem}
\emph{Proof:} $(i)$ \ \ Let $f : M \to \RR$ be continuous. If $\gl < \mu$
then clearly $\ef_{\gl} \tm \ef_{\mu}$ and therefore $\ef_{\gl} \tm
\We_{\mu > \gl}\ef_{\mu}$. On the other hand
\begin{eqnarray*}
    int (\bigcap_{\mu > \gl}\ef_{\mu} & = & int (\bigcap_{\mu > \gl}int
    (\overset{-1}{f}(]-\∞, \mu])))  \\
    & \tm & int (\bigcap_{\mu > \gl}\overset{-1}{f}(]-\∞, \mu]))  \\
    & = & int (\overset{-1}{f}(\bigcap_{\mu > \gl}]-\∞, \mu]))  \\
    & = & int (\overset{-1}{f}(]-\∞, \gl]))  \\
    & = & \ef_{\gl},
\end{eqnarray*}
hence $\ef_{\gl} = \We_{\mu > \gl}\ef_{\mu}$. \\
Assume that there is some $x \in \We_{\gl \in \RR}\ef_{\gl}$. Then $x \in
\bigcap_{\gl \in \RR}\overset{-1}{f}(]-\∞, \gl])$, i.e. $f(x) ≤ \gl$
for all $\gl \in \RR$, which is absurd. Hence $\bigcap_{\gl \in
\RR}\ef_{\gl} = \emptyset$ and therefore
\[
    \We_{\gl \in \RR}\ef_{\gl} = \emptyset \quad \text{and} \quad
    \kD(\ef) = M.
\] 
As $f$ is continuous we have $\overset{–1}{f}(]-\∞, \gl[) \tm \ef_{\gl}$ and
therefore $\bigcup_{\gl \in \RR}\ef_{\gl} = M$. Hence $\ef$ is a
spectral family. Note that we have used the continuity of $f$ only for
the property $\bigcup_{\gl \in \RR}\ef_{\gl} = M$. \\
Eventually we prove that $\ef$ is strongly regular. We show first that
$\gl < \mu$ implies $(\overset{-1}{f}(]\mu, \∞[))^{-} \tm \overset{-1}{f}
(]\mu, \∞[)$. Indeed, assume that $x \in (\overset{-1}{f}(]\mu, \∞[))^{-}$
but $f(x) ≤ \gl$. Let $U$ be an open neighborhood of $x$ such that
$f(y) < \frac{\gl + \mu}{2}$ for all $y \in U$. Because of $U \cap
\overset{-1}{f}(]\mu, \∞[) \neq \emptyset$ there is some $y_{0} \in U$
such that $f(y_{0}) > \mu$. But this gives the contradiction $\mu <
f(y_{0}) <  \frac{\gl + \mu}{2} < \mu$. \\
Taking complements we can express this result in the following way:
\[
    \mu > \gl \quad \lra \quad \ef_{\gl} \tm \overset{-1}{f}(]-\∞, \gl])
    \tm \ef_{\mu}.
\]
$\overset{-1}{f}(]-\∞, \gl])$ is closed because $f$ is continuous and 
therefore
\[
    \overline{\ef_{\gl}} \tm \overset{-1}{f}(]-\∞, \gl]) \tm \ef_{\mu}.
\]
~\\
$(ii)$ \ \ Now assume that $E : \RR \to \kT(M)$ is a
\emph{strongly regular } spectral family. We show that the induced function
$f_{E} : \kD(E) \to \RR$ is continuous. It suffices to prove that 
$\overset{-1}{f_{E}}(]\gl, \mu[)$ is open for all $\gl < \mu$. Let
$x$ be an element of this set (remember that $\kD(E)$ is an open and
dense subset of $M$). Then there is some $\eps > 0$ such that $\gl +
2\eps < f_{E}(x) < \mu - \eps$. As $E$ is strongly regular  we have
$\overline{E_{\gl + \eps}} \tm E_{\gl + 2\eps}$. From the
definition of $f_{E}$ we conclude that $x$ belongs to $E_{\mu -
\eps} \smm \overline{E_{\gl + \eps}}$ and that $f_{E}$ maps the
open neighborhood  $E_{\mu - \eps} \smm \overline{E_{\gl + \eps}}
\cap \kD(E)$ of $x$ into $]\gl, \mu[$. Hence $f_{E}$ is continuous.
\\
~\\
$(iii)$ \ \ Let $f : M \to \RR$ be a continuous function. We show that
\[
    f_{\ef} = f
\]
holds.  Let $x \in M$ and $\gl_{0} := f_{\ef}(x) = \inf \{ \gl \ |
\ x \in \ef_{\gl} \}$.  If $x \in \ef_{\gl}$ then $f(x) ≤ \gl$ and therefore
$f(x) ≤ \gl_{0}$.  Assume that $f(x) < \gl_{0}$.  Then $f(x) < \gl_{0}
- \eps$ for some $\eps > 0$ and $x \in int (\overset{-1}{f}(]-\∞,
\gl_{0} - \eps]))$ because $\overset{-1}{f}(]-\∞, \gl_{0} - \eps[)$ is
an open subset of $\overset{-1}{f}(]-\∞, \gl_{0} - \eps])$.  Therefore
$x \in \ef_{\gl_{0} - \eps}$ contrary to the definition of
$f_{\ef}(x)$. Hence $f_{\ef}(x) = f(x)$. \\
~\\
$(iv)$  We finally show that for all strongly regular spectral
families $E$
\[
    \forall \gl \in \RR : \ E^{f_{E}}_{\gl} = \el \cap \kD(E)
\]
holds. \\ Let $x \in \el \cap \kD(E)$. Then $f_{E}(x) ≤ \gl$ and
therefore 
\[
     \el \cap \kD(E) \tm \overset{-1}{f_{E}}(]-\∞, \gl]).
\]     
Hence
\[
     \el \cap \kD(E) \tm E^{f_{E}}_{\gl}
\]
since $\el$ is open. Conversely, let $x \in E^{f_{E}}_{\gl}$ and
let $U$ be an open neighborhood of $x$ contained in
$\overset{-1}{f_{E}}(]-\∞, \gl])$. Let $\gl_{0} := f_{E}(x)$ and
let $]\gl_{0} - \eps, \gl_{0} + \eps[$ be an open interval around
$\gl_{0}$ such that $\overset{-1}{f_{E}}(]\gl_{0} - \eps, \gl_{0} +
\eps[)$ is contained in $U$. In particular we obtain 
\[
    \overset{-1}{f_{E}}(]\gl_{0} - \eps, \gl_{0} + \eps[) \tm 
    \overset{-1}{f_{E}}(]-\∞, \gl]).
\] 
Assume that $\gl_{0} = \gl$. Then there is no $y \in \kD(E)$ such
that $\gl < f_{E}(y) < \gl + \eps$ and therefore $E$ is constant
on the interval $]\gl, \gl + \eps[$. Hence
\[
    \el = \We_{\mu > \gl}\emm = int (\bigcap_{\gl < \mu < \gl +
    \eps}\emm) = \emm.
\]
Therefore $E$ is constant on the interval $[\gl, \gl + \eps[$ and
this shows $x \in \el$. If $\gl_{0} < \gl$ then clearly $x \in \el$ 
and $\el \cap \kD(E) = \overset{-1}{f_{E}}(]-\∞, \gl])$ is proved.
\ \ $\Box$ 

\begin{remark}\label{co34}
    The foregoing proof has shown that for an \emph{arbitrary}
    function $f : M \to \RR$ the corresponding map $\ef : \RR \to 
    \kT(M)$ fails to be a spectral family only in one point: the
    property $\bigcup_{\lir}\ef_{\gl} = M$ must not be satisfied. This is
    shown by the following simple example: let $f : [0, \∞[ \to \RR$
    be defined by $f(0) = 0$ and $f(x) = \frac{1}{x}$ for $x > 0$.
    Then $0$ is contained in $\bigcap_{\gl}(\overset{-1}{f}(]\gl,
    \∞[))^{-}$, the complement of $\bigcup_{\gl}\sfl$. \\
    Of course, this phenomenon cannot occur if $f$ is bounded from
    above. So in particular every \emph{bounded} function $f : M \to
    \RR$ induces a spectral family $\ef$.\\
    ~\\
    If $f : M \to \RR$ is a continuous function, then the natural
    guess for defining a corresponding spectral family would be
    \[
	\gl \tto \urb{f}(]-\∞, \gl [).
    \]
    In general, this is only a \emph{pre-spectral family}: it
    satisfies all properties of a spectral family, except continuity
    from the right. This is cured by \emph{spectralization}, i.e. by
    the switch to
    \[
	\gl \tto \We_{\mu > \gl}\urb{f}(]-\∞, \mu [).
    \]
    But
    \[
    \We_{\mu > \gl}\urb{f}(]-\∞, \mu [) = int (\bigcap_{\mu > \gl}
    \urb{f}(]-\∞, \mu [)) = int (\urb{f}(]-\∞, \gl ])),
    \]
    which shows that our original definition is the natural one.
\end{remark}
		       
\begin{remark}\label{co35}
    Note that the spectral family $\ef$ of a continuous function
    $f : M \to \RR$ is strongly regular but not necessarily continuous in the
    usual sense. This is shown by the following example: Let $M$ be
    disconnected and let $M_{0}$ be a nonvoid open and closed subset
    different from $M$. Then the spectral family of $f :=
    \chi_{M_{0}}$ is continuous from the right but not from the left.
\end{remark}
~\\
According to theorem \ref{co33} $f \tto \ef$ is a one-to-one
correspondence between bounded continuous functions $f : M \to \RR$
and bounded \emph{strongly regular} spectral families $\ef$ in $\ktm$. A
strongly regular  spectral family takes its values in the lattice $\ktrm$, and
we know that $\ktrm$ is a \emph{complete Boolean algebra} with respect
to the operations
\begin{enumerate}
    \item  [(i)] $\Ve_{\kik}U_{\gk} := (\bigcup_{\kik}U_{\gk})^{cc}$, 

    \item  [(ii)] $\We_{\kik}U_{\gk} := int \bigcap_{\kik}U_{\gk}$,

    \item  [(iii)] $U \tto U^c := M \smm \overline{U}$.
\end{enumerate}
In particular, we can regard $\ktrm$ as a $\gs$-algebra. So a
continuous function $f : M \to \RR$ is a generalized $\ktrm$-measurable
function in the sense of definition \ref{co5}. Hence we can apply the 
results of the previous section to bounded continuous functions $f : M
\to \RR$. This is of particular interest if $M$ is a \emph{Baire
space}, because in this case $\ktrm$ is $\gs$-isomorphic to the
quotient $\kbm / \kI_{1}$, where $\kbm$ is the $\gs$-algebra of all
Borel subsets of $M$ and $\kI_{1}$ the $\gs$-ideal of all meagre Borel
sets.

\begin{remark}\label{co36}
    We call a spectral family $E : \RR \to \kbm$ regular if $\el \in
    \ktrm$ for all $\lir$. Although $\ktrm$ is not a sublattice of
    $\kbm$, we can regard a regular spectral family $E$ as a spectral
    family in $\ktrm$: Since $\el = \bigcap_{\mu > \gl}\emm$ in the
    lattice $\kbm$ and $\el$ is open, $\el = \We_{\mu > \gl}\emm$ in
    the lattice $\ktrm$. Note that a regular spectral family is not
    necessarily strongly regular. This is shown by the following
    example. Let $f : M \to \RR$ be a bounded lower semicontinuous
    function that is not continuous. Since $\urb{f}(]-\∞, \gl])$ is
    closed for all $\lir$ (this property characterizes lower
    semicontinuity) and the interior of a closed set is regular, the
    spectral family $E^f$ of $f$ is regular, but not strongly regular.    
\end{remark}
Let $f : M \to \RR$ be a continuous function. For $\lir$ let
\[
    E^f_{\gl} := int \urb{f}(]-\∞, \gl]) \quad \text{and} \quad
    F^f_{\gl} := \urb{f}(]-\∞, \gl]).
\]
Let $M$ be a Baire space. Since $\urb{f}(]-\∞, \gl])$ is closed,
$int \urb{f}(]-\∞, \gl])$ is the unique regular representative of
the equivalence class of $\urb{f}(]-\∞, \gl])$ modulo the $\gs$-ideal 
$\kI_{1}$ of all meagre Borel subsets of $M$. Hence the spectral family
of a continuous function $f : M \to \RR$ in $\ktrm$ is the quotient of
the spectral family of $f$, regarded a Borel-measurable function on 
$M$, in $\kbm$ modulo $\kI_{1}$.\\
~\\
The function $f_{E}$ induced by the spectral family $E$ is
defined on a dense subset of the topological space $M$. Of
course $E$ induces a function on (a subset of)
the set $\kD(\kT(M))$ of dual ideals of $\kT(M)$ and in particular
on (a subset of) the Stone spectrum $\kQ(M)$ of $\kT(M)$ in
the usual manner:
\[  f_{E}(\kj) := \inf \{ \gl | \el \in \kj \},  \]
which is defined on the set $\kD_{ad}(\ktm)$ of all dual ideals $\kj$
that satisfy
\[
\emptyset \ne \kj \cap im(E) \ne im(E).
\]
This condition is satisfied in particular for {\bf bounded}
spectral families.\\
$f_{E} : \kD_{ad}(\kT(M)) \to \RR$ is an extension of $f_{E}
: \kD(E) \to \RR$ because $x \in \el$ if and only if
$\el \in \frp_{x}$ where $\frp_{x} \in \kD(\kT(M))$ is the 
point defined by $x \in M$.

\begin{definition}\label{co37}
    A quasipoint $\frb$ in $\ktm$ is called a quasipoint over $x \in
    M$, if $x \in \bigcap_{U \in \frb}\overline{U}$. We denote the set
    of quasipoints over $x$ by $\kQ_{x}(\ktm)$.
\end{definition}

\begin{remark}\label{co38}
    If $M$ is locally compact, $\frb \in \qm$ is a quasipoint over
    some $x \in M$ if and only if $\frb$ is of finite type. 
\end{remark}

\begin{proposition}\label{co39}
    Let $M$ be a Hausdorff space. Then $\qxm \ne \emptyset$ for all $x
    \in M$. Moreover, $\bigcup_{x \in M}\qxm$ is dense in $\qm$.
\end{proposition}
\emph{Proof:} The set of open neighbourhoods of $x \in M$ is a filter 
base, hence it is contained in some quasipoint $\frb \in \qm$. If $U
\frb$, then $U \cap V \ne \emptyset$ for all open neighbourhoods $V$
of $x$, so $x \in \overline{U}$.\\
Let $U \in \ktm$ and let $x \in U$. Take any quasipoint $\frb$ that
contains all open neighbourhoods of $x$. Then $\frb \in \qxm \cap
\kQ_{U}(\ktm)$, hence $\bigcup_{x \in M}\qxm$ is dense in $\qm$. \ \
$\Box$ \\
~\\
Since $M$ is Hausdorff, $\bigcap_{U \in \frb}\overline{U}$ contains at
most one element for every $\frb \in \qm$. Let $\frp_{x}$ be the set
of all open neighbourhoods of $x \in M$, i.e. the point in $\ktm$
defined by $x$. It is now obvious that $\frb \in \qxm$ if and only if 
$\frp_{x} \tm \frb$. Let
\[
    \kQ^{pt}(\ktm) := \bigcup_{x \in M}\qxm.
\]
We obtain a mapping 
\[
    pt : \kQ^{pt}(\ktm) \to M,
\]
defined by $pt(\qxm) := \{x\}$.

\begin{proposition}\label{co40}
    The topology of $M$ is the identification topology with respect to
    $pt : \kQ^{pt}(\ktm) \to M$.
\end{proposition}
\emph{Proof:} Let $U$ be an open subset of $M$. Then 
\[
    \urb{pt}(U) = \bigcup_{x \in U}\qxm =  \kQ^{pt}(\ktm) \cap
    \kQ_{U}(\ktm),
\]
which is open in $\kQ^{pt}(\ktm)$. Hence $pt$ is continuous and,
therefore, the identification topology is finer then the given topology
of $M$.\\
Let $X \tm M$ such that $\urb{pt}(X)$ is open in
$\kQ^{pt}(\ktm)$. If $x \in X$ and $\frb$ is a quasipoint over $x$,
then there is some $U \in \ktm$ such that $\frb \in \kQ^{pt}(\ktm) 
\cap \kQ_{U}(\ktm) \tm \urb{pt}(W)$. The difficulty is, that we can
only conclude that $x$ belongs to $\overline{U}$!\\
Assume that $X$ is not open and let $x \in X \smm int X$. Consider the
subset
\[
    C_{x} := \{ U^c \mid U \in \ktm, \ x \in \overline{U} \tm X \}
\]
of $\ktm$. We show that $C_{x}$ is a filter base in $\ktm$. It is obvious
that $\emptyset \notin C_{x}$. Let $U^c, V^c \in C_{x}$. Then $U^c
\cap V^c = (U \cup V)^c$, and $x \in \overline{U}, \overline{V} \tm X$
implies $x \in \overline{U} \cup \overline{V} = \overline{U \cup V}
\tm X$, hence $U^c \cap V^c \in C_{x}$. In the next step we show that 
$x \in \overline{U^c}$ for all $U^c \in C_{x}$. Assume that this is
not the case. Then $x \in M \smm \overline{V^c} = int \overline{V}$
for some $V^c \in C_{x}$, but $\overline{V} \tm X$ implies $x \in int 
X$, a contradiction to the choice of $x$. Thus $W \cap U^c \ne
\emptyset$ for all open neighbourhoods $W$ of $x$ and all $U^c \in
C_{x}$. Altogether we have proved that there is a quasipoint $\frb$
over $x$ that contains $C_{x}$. Since $\urb{pt}(X)$ is open in 
$\kQ^{pt}(\ktm)$, there is some $U \in \ktm$ such that $\kQ^{pt}(\ktm) 
\cap \kQ_{U}(\ktm) \tm \urb{pt}(X)$. Then $x \in \overline{U} \tm X$
and, according to the construction of $\frb$, $U^c \in \frb$. But this
gives the contradiction $\emptyset = U \cap U^c \in \frb$. \ \ $\Box$

\begin{corollary}\label{co41}
    $\kQ^{pt}(\ktm) =\qm$ if and only if $M$ is compact. 
\end{corollary}

\begin{proposition}\label{co42}
    Let $E : \RR \to \kT(M)$ be a strongly regular spectral family and let
    $x \in \kD(E)$. Then for all quasipoints $\frb_{x} \in \qm$ over
    $x$ we have
    \[
    f_{E}(\frb_{x}) = f_{E}(x).
    \]
\end{proposition} 
\emph{Proof:} Because of $f_{E}(x) = f_{E}(\frp_{x})$ we have 
\[
f_{E}(\frb_{x}) ≤ f_{E}(x).
\]
For $\eps > 0$ we can choose $\gl \in \RR$ such that $\gl <
f_{E}(\frb_{x}) + \eps$ and $\el \in \frb_{x}$. Then $x \in
\overline{\el}$ and for $\gl < \mu < f_{E}(\frb_{x})$ we obtain
from the strong regularity of $E$ that $x \in \overline{\el} \tm
\emm \in \frb_{x}$ holds. Thus
\[
f_{E}(x) ≤ \mu < f_{E}(\frb_{x}) + \eps 
\]
and therefore $f_{E}(x) ≤ f_{E}(\frb_{x})$.  \ \ $\Box$ \\
~\\
Let $f : M \to \RR$ be a \emph{bounded} continuous function with spectral
family $E$. Then the associated function $f_{E} : \qm \to \RR$ is continuous.
This follows from the proof of proposition \ref{co8}, using the fact
that $E$ is a regular spectral family. The restriction of $f_{E}$ to
$\kQ^{pt}(\ktm)$ is, according to proposition \ref{co42}, constant on 
the fibres of $pt : \kQ^{pt}(\ktm) \to M$.\\
Conversely, let $\gf : \qm \to \RR$ be any continuous function
such that $\gf\mid_{\kQ^{pt}(\ktm)}$ is constant on the fibres of $pt$.
Since $\qm$ is compact, $\gf$ is bounded. Then $\gf$ induces a unique
bounded function $f_{\gf} : M \to \RR$ such that 
\[
    f_{\gf} \circ pt = \gf\mid_{\kQ^{pt}(\ktm)}.
\]
Due to proposition \ref{co40}, $pt$ is identifying, hence $f_{\gf}$ is
a bounded continuous function on $M$. Let $\egf$ be the spectral
family corresponding to $\egf$. It follows from proposition \ref{co42}
that $(f_{\egf})\mid_{\kQ^{pt}(\ktm)} = \gf\mid_{\kQ^{pt}(\ktm)}$.
Since $\kQ^{pt}(\ktm)$ is dense in $\qm$, this implies $f_{\egf} =
\gf$. Moreover, this also shows that a bounded continuous function
$\psi : \kQ^{pt}(\ktm) \to \RR$, which is constant on each fibre of
$pt$, can be extended to a continuous function $\gf : \qm \to \RR$. \\
~\\
Let $C^{pt}(\qm)$ be the set of all continuous functions $\qm \to \CC$
that are constant on each fibre of $pt$. Clearly, $C^{pt}(\qm)$ is a
$C^{\ast}$-subalgebra (with unity) of $C(\qm)$. Since $C^{pt}(\qm)$ is
selfadjoint, the foregoing considerations apply to complex valued functions
in $C^{pt}(\qm)$ as well. It is now easy to see that the mapping
\[
    \begin{array}{cccc}
        f_{\ast} : & C_{b}(M) & \to & C^{pt}(\qm)  \\
         & \gf & \tto & f_{E^{\Re \gf}} + i f_{E^{\Im \gf}},
    \end{array}
\]
where $\Re \gf$ and $\Im \gf$ denote the real and imaginary part of
$\gf$ respectively, is a $\ast$-isomorphism of $C^{\ast}$-algebras. We
can express $f_{\ast}(\gf)$ also as $f_{\ast}(\gf) = f_{E^{\gf}}$ where
$E^{\gf}$ is the \emph{complex} spectral family corresponding to $\gf$.\\
~\\
By theorem \ref{co23}, the set $\kS_{b}(\CC, \ktrm)$ of all bounded
\emph{regular} complex spectral families is a $C^{\ast}$-algebra which
is $\ast$-isomorphic to $C(\qrm)$. $C(\qrm)$ is $\ast$-isomorphic to
$C(\qm)$, since $\qrm$ is homeomorphic to $\qm$. The foregoing results
show how the continuous functions $\qm \to \CC$ that are induced by 
bounded strongly regular complex spectral families in $\ktm$, or
equivalently by bounded continuous functions $M \to \CC$, can be
characterized in $C(\qm)$. 
 
\begin{theorem}\label{co43}
    Let $M$ be a Hausdorff space, $C_{b}(M)$ the $C^{\ast}$-algebra of
    all bounded continuous functions $M \to \CC$ and $C^{pt}(\qm)$ the
    $C^{\ast}$-subalgebra of $C(\qm)$, consisting of all $\gf$ in
    $C(\qm)$ that are constant on each fibre of $pt : \kQ^{pt}(\ktm)
    \to M$. Then the mapping
    \[
	\begin{array}{cccc}
	    f_{\ast} : & C_{b}(M) & \to & C^{pt}(\qm)  \\
	     & \gf & \tto & f_{E^{\gf}},
	\end{array}
    \]
    where $E^{\gf}$ is the complex spectral family corresponding to
    $\gf$, is a $\ast$-isomorphism of $C^{\ast}$-algebras.
\end{theorem}

\begin{corollary}\label{co44}
    Let $M$ be a completely regular Hausdorff space. Then the
    Stone-\v{C}ech compactification $\check{M}$ of $M$ is homeomorphic to 
    the Gelfand spectrum $\gG(C^{pt}(\qm))$ of the abelian
    $C^{\ast}$-algebra $C^{pt}(\qm)$. 
    If, in particular, $M$ is discrete, $\check{M}$ is homeomorphic to
    $\qm$. 
\end{corollary}
\emph{Proof:} The first assertion follows from theorem \ref{co43} and 
the well known fact that $\check{M}$ is homeomorphic to the Gelfand
spectrum of $C_{b}(M)$. If $M$ is discrete, $\qxm$ consists of exactly
one element, namely $\frb_{x} := \{ U \tm M \mid x \in U \}$, hence
$C^{pt}(\qm) = C(\qm)$. \ \ $\Box$\\ 
~\\
Finally we prove, in analogy to the the case of measurable functions, 
a {\bf spectral representation for continuous functions.}\\
~\\
Let $M$ be a Hausdorff space. Continuous functions $\gf : M \to \RR$
are characterized by strongly regular spectral families $\egf$ with
values in $\ktrm$ and admissible domain $M$.\\
Similar to the case of measurable functions we consider for $\eps > 0$
a sequence $(\gl_{k})_{k \in \ZZ}$ of real numbers with $\gl_{k + 1} -
\gl_{k} = \eps$ for all $k \in \ZZ$. The use of the differences
$\egf_{\gl_{k + 1}} \smm \overline{\egf_{\gl_{k}}}$, although
natural\footnote{Since $\overline{\egfl} \tm \egf_{\mu}$ for all $\mu > \gl$,
$\egf_{\mu} \smm \overline{\egfl}$ is just the $\ktrm$-complement of
$\egfl$ in $\egf_{\mu}$: $\all \ \mu > \gl : \ \egf_{\mu} \we (\egfl)^{\pp}
= \egf_{\mu} \smm \overline{\egfl}$.}, would lead to some messy
technical problems. Moreover, their characteristic functions are, in
general, only Borel-measurable and not continuous. Therefore we use the
spectral family $B^{\gf}$ of $\gf$, considered as a Borel-measurable
function:
\[
    \all \ \lir : \ B^{\gf}_{\gl} := \urb{\gf}(]- \∞, \gl]).
\] 
Then
\[
    \gf = \int_{\RR}\gl d\chi_{B^{\gf}_{\gl}}. 
\]
~\\
If the continuous function $\gf : M \to \RR$ is bounded, we obtain a
natural spectral representation for $\gf$ via the spectral representation
for the associated function $f_{\egf} : \qm \to \RR$. $\egf$ is a
bounded generalized $\ktrm$-measurable function. We know from the
previous section that
\[
    \all \ \frb \in \kQ(\ktrm) : \ f_{\egf}(\frb) = \int_{\RR}\gl
    d\chi^{\kQ}_{\egfl}(\frb).
\]
Thus $f_{\egf}(\frb)$ is a Riemann-Stieltjes integral with respect to 
the bounded increasing function $\gl \tto
\chi_{\kQ_{\egfl}(\ktrm)}(\frb)$. Since $\gf \circ pt = f_{\egf}
\mid_{\kQ^{pt}(\ktrm)}$, we obtain 
\[
    \all \ x \in M : \ \gf(x) = \int_{\RR}\gl
    d\chi^{\kQ}_{\egfl}(\frb_{x}),
\]
where $\frb_{x}$ is any quasipoint over $x$. This is the spectral
representation for bounded continuous functions.

\section{Common Structure of Quantum and Classical Observables}
\label{qco}
\pagestyle{myheadings}
\markboth{Classical Observables}{Common Structure of Quantum and
Classical Observables}

Let $M$ be a Hausdorff space, $\ktrm_{0} := \ktrm \smm \{\emptyset\}$,
and let $E$ be a bounded spectral family in $\ktrm$ with associated function
$f_{E} \in C(\qm)$. Since
\[
    \all \ U \in \ktrm_{0} : \ H_{U} = \bigcap_{\frb \in
    \kQ_{U}(\ktrm)}\frb,
\]
we can define 
\[
    r_{E}(U) := \sup \{ f_{E}(\frb) \mid U \in \frb \}.
\]
As in the case of operator algebras, $r_{E} : \ktrm_{0} \to \RR$ is
completely increasing:
\[
    r_{E}(\Ve_{\kik}U_{k}) = \sup_{\kik}r_{E}(U_{k})
\]
for all families $(U_{k})_{\kik}$ in $\ktrm_{0}$. Starting from an
arbitrary function $f \in C(\qrm, \RR)$, the same construction gives a
function $r_{f} : \ktrm_{0} \to \RR$. $r_{f}$ is completely
increasing, too. The proof rests on the following %take this into a new
%version of part I%
\begin{lemma}\label{co45}
    If $(U_{k})_{\kik}$ is an arbitrary family in $\ktrm$, then 
    \[
    \overline{\bigcup_{\kik}\kQ_{U_{k}}(\ktrm)} =
    \kQ_{\Ve_{\kik}U_{k}}(\ktrm).
    \]
    Consequently, $\ktrm$ is a completely distributive lattice.
\end{lemma}
\emph{Proof:} Since $\ktm$ is a completely distributive lattice, the
Stone spectrum $\qm$ is a Stonean space, i.e. the closure of every
open set is open. Hence also $\qrm$ is a Stonean space, because it is 
homeomorphic to $\qm$. As $\ktrm$ is a distributive ortholattice, the 
open closed subsets of $\qrm$ are of the form $\kQ_{V}(\ktrm)$. We
therefore conclude that for every family $(U_{k})_{\kik}$ in $\ktrm$
there is a $V \in \ktrm$ such that   
\[
    \overline{\bigcup_{\kik}\kQ_{U_{k}}(\ktrm)} = \kQ_{V}(\ktrm).
\]  
It follows that $\kQ_{U_{k}}(\ktrm) \tm \kQ_{V}(\ktrm)$ for all
$\kik$, hence $U_{k} \tm V$ for all $\kik$. Therefore $\Ve_{\kik}U_{k}
\tm V$. On the other hand, since $\kQ_{\Ve_{\kik}U_{k}}(\ktrm)$ is
also closed, we have $\overline{\bigcup_{\kik}\kQ_{U_{k}}(\ktrm)} \tm 
\kQ_{\Ve_{\kik}U_{k}}(\ktrm)$, thus $V = \Ve_{\kik}U_{k}$. This
implies, by corollary 3.1 in \cite{deg3}, that $\ktrm$ is completely
distributive. \ \ $\Box$\\
~\\
Now the same argument as in \cite{deg4} for abelian von Neumann algebras
shows

\begin{remark}\label{co46}
    For every $f \in C(\qrm, \RR)$, the induced function
    \[
        \begin{array}{cccc}
            r_{f} : & \ktrm_{0} & \to & \RR  \\
             & U & \tto & \sup_{\frb \in \kQ_{U}(\ktrm)}f(\frb)
        \end{array}
    \]
    is completely increasing.
\end{remark}
We denote by $\frp^{r}_{x}$ the set of all regular open neighbourhoods of
$x$. Obviously, $\frb \in \qrm$ is a quasipoint over $x$ if and only
if $\frp^{r}_{x} \tm \frb$. Let $U \in \ktrm$ be a neighbourhood of $x \in M$
and let $V \in \ktrm$. $U \cap V^{c} = \emptyset$ implies $U = (U \cap V)
\vee (U \cap V^{c}) = U \cap V$, hence $V$ contains $U$, i.e. is a neighbourhood
of $x$. Therefore, if $\frb \in \qxrm$ and if $V \in \frb$ is not a neighbourhood
of $x$, then there is a quasipoint over $x$ that contains $V^{c}$.
This shows that
\[
     \all \ x \in M : \ \frp^{r}_{x} = \bigcap_{\frb \in \qxrm}\frb
\]
holds. More generally, it is not difficult to show that 
\[
    \all \ x \in M : \ \bigcap_{\frb \in \qxm}\frb = \{ U \in \ktm
    \mid U^{cc} \in \frp^{r}_{x} \}
\]
holds in the lattice $\ktm$. Note that $\{ U \in \ktm \mid U^{cc} \in \frp^{r}_{x}
\}$ is, in general, strictly larger than $\frp_{x}$. \\
~\\
We know from theorem \ref{co43}, that the continuous functions $f : \qrm
\to \RR$ that are induced by bounded continuous functions $\gf : M \to \RR$ 
are precisely those that are constant on each fibre of $pt : \kQ^{pt}(\ktrm) \to M$.
This property can be formulated in terms of the corresponding completely
increasing function $r_{f} : \ktrm_{0} \to \RR$ as
\[
     (\ast) \qquad \all \ x \in M \ \all \ \frb \in \qxrm : \ \inf_{U \in
     \frp^{r}_{x}}r_{f}(U) = \inf_{V \in \frb}r_{f}(V).
\]
~\\
The {\bf main open problem} for classical observables is a structural 
characterization of those $f \in C_{b}(\qrm, \RR)$ that are associated
to bounded {\bf smooth functions on a smooth manifold $M$}. \\

\begin{discussion} 
We have seen in this and the previous part that
bounded quantum as well as bounded classical observables can be 
represented as bounded {\bf continuous functions} on the Stone
spectrum of the corresponding lattice. A dual common feature is that
both quantum and classical observables can be described by
{\bf spectral families} in the corresponding lattice. The link between
these two descriptions is the assignment of a function
$f_{E} \in C_{b}(\ql, \RR)$ to every bounded spectral family $E$ in an
orthomodular lattice $\LL$. \\
~\\
For the commutative case, i.e. for selfadjoint elements
of an abelian von Neumann algebra, for real valued (generalized)
measurable functions and for real valued continuous functions, the
mapping $E \to f_{E}$ is essentially the Gelfand transformation. This 
follows from theorem 2.9 in \cite{deg4} and theorems
\ref{co14}, \ref{co43}. Therefore, we see that measure theory is, in
a definite sense, a generalization of the theory of abelian von Neumann
algebras: the $C^{\ast}$-algebra of all bounded generalized
$\frA$-measurable functions is an abelian von Neumann algebra if and
only if the Stone spectrum $\qA$ of the $\gs$-algebra $\frA$ is a
hyperstonean space. (See \cite{tak1} for the discussion of
hyperstonean spaces.) \\
~\\
If the Stone spectrum $\qA$ is not hyperstonean but only a Stonean
space, we get the measure theory of {\bf completely distributive}
Boolean algebras. Note that a completely distributive Boolean algebra 
is an ortholattice $\LL$ with the following two properties:
\begin{enumerate}
    \item  [(i)] $\LL$ is distributive,

    \item  [(ii)] $\overline{\bigcup_{\kik}\kQ_{a_{k}}(\LL)} =
    \kQ_{\Ve_{\kik}a_{k}}(\LL)$ for every increasing family
    $(a_{k})_{\kik}$ in $\LL$.
\end{enumerate}
Now replace property $(i)$ by
\begin{enumerate}
    \item  [(i')] $\LL$ is complete and orthomodular.
\end{enumerate}
Of course, each finite orthomodular lattice satisfies conditions
$(i')$ and $(ii)$. If $\rr$ is a von Neumann algebra, the projection
lattice $\pr$ of $\rr$ satisfies $(i'), (ii)$ if and only if $\rr$ is 
finite (\cite{deg3}, theorem 1.3).\\
~\\
Bounded quantum observables belonging to a fixed ``context'', i.e. to a fixed
(maximal) abelian von Neumann subalgebra $\mm$ of some von Neumann
algebra $\rr$, and bounded (generalized) $\frA$-measurable functions
have in common that the corresponding maps $f_{\ast} : E \tto f_{E}$ to
$C(\kQ(\mm), \RR)$ and $C(\qA, \RR)$, respectively, are surjective. If
the von Neumann algebra $\rr$ is not abelian, $f_{\ast}$ is not
surjective, and only those $f \in C_{b}(\qr, \RR)$ are in the range of
$f_{\ast}$ for which the induced function $r_{f} : \por \to \RR$ is
completely increasing.\\
~\\
Altogether we have shown that both quantum and classical observables
can be described either by  spectral families in an associated lattice
(depending on the context) or by continuous real valued functions on
the Stone spectrum of the associated lattice. The link between these
different descriptions is given by the mapping $f_{\ast}$, a canonical
generalization of the Gelfand transformation. Moreover, we have shown 
that $f_{\ast}$ coincides (essentially) with the Gelfand transformation
if and only if the corresponding algebra of observables is abelian.
\end{discussion}

\end{document}